\documentclass[manuscript]{acmart}

\AtBeginDocument{%
  }

\copyrightyear{2024}
\acmYear{2024}
\setcopyright{acmlicensed}\acmConference[MUM '24]{International Conference on Mobile and Ubiquitous Multimedia}{December 1--4, 2024}{Stockholm, Sweden}
\acmBooktitle{International Conference on Mobile and Ubiquitous Multimedia (MUM '24), December 1--4, 2024, Stockholm, Sweden}
\acmDOI{10.1145/3701571.3701593}
\acmISBN{979-8-4007-1283-8/24/12}

\acmISBN{978-1-4503-XXXX-X/18/06}

\usepackage{multirow}

\usepackage{booktabs,caption}

\usepackage{rotating}
\usepackage{geometry}

\usepackage{soul}

\usepackage{graphicx}
\usepackage{array}

\usepackage{listings}
\usepackage{color}

\usepackage{listings}
\usepackage{xcolor}
\usepackage{hyperref}
\usepackage{todonotes}

\usepackage{amsmath} 

\usepackage{booktabs}

\usepackage{framed}

\definecolor{codegreen}{rgb}{0,0.6,0}
\definecolor{codegray}{rgb}{0.5,0.5,0.5}
\definecolor{codepurple}{rgb}{0.58,0,0.82}
\definecolor{backcolour}{rgb}{0.96,0.97,0.96}

\lstdefinestyle{prompts}{
  backgroundcolor=\color{backcolour}, commentstyle=\color{codegreen},
  keywordstyle=\color{magenta},
  numberstyle=\tiny\color{codegray},
  stringstyle=\color{codepurple},
  basicstyle=\ttfamily\scriptsize,
  breakatwhitespace=false,         
  breaklines=true,                 
  captionpos=b,                    
  keepspaces=false,                 
  numbers=left,                    
  numbersep=5pt,                  
  showspaces=false,                
  showstringspaces=false,
  showtabs=false,                  
  tabsize=2
}

\begin{document}

\title{From Voice to Value: Leveraging AI to Enhance Spoken Online Reviews on the Go}

\author{Kavindu Perera}
\email{kavindu.magalage@oulu.fi}
\orcid{0000-0001-9621-3104}
\affiliation{%
  \institution{University of Oulu}
  \city{Oulu}
  \country{Finland}
}

\author{Dániel Szabó}
\orcid{0009-0003-7299-9116}
\email{daniel.szabo@oulu.fi}
\affiliation{%
  \institution{University of Oulu}
  \city{Oulu}
  \country{Finland}
}

\author{Niels van Berkel}
\orcid{0000-0001-5106-7692}
\email{nielsvanberkel@cs.aau.dk}
\affiliation{%
  \institution{Aalborg University}
  \city{Aalborg}
  \country{Denmark}
}

\author{Aku Visuri}
\email{aku.visuri@oulu.fi}
\orcid{0000-0001-7127-4031}
\affiliation{%
  \institution{University of Oulu}
  \city{Oulu}
  \country{Finland}
}

\author{Chi-Lan Yang}
\email{chilan.yang@iii.u-tokyo.ac.jp}
\orcid{0000-0003-0603-2807}
\affiliation{%
  \institution{The University of Tokyo}
  \city{Tokyo}
  \country{Japan}
}

\author{Koji Yatani}
\email{koji@iis-lab.org}
\orcid{0000-0003-4192-0420}
\affiliation{%
  \institution{The University of Tokyo}
  \city{Tokyo}
  \country{Japan}
}

\author{Simo Hosio}
\email{simo.hosio@oulu.fi}
\orcid{0000-0002-9609-0965}
\affiliation{%
  \institution{University of Oulu}
  \city{Oulu}
  \country{Finland}
}

\renewcommand{\shortauthors}{Perera et al.}

\begin{abstract}
Online reviews help people make better decisions. Review platforms usually depend on typed input, where leaving a good review requires significant effort because users must carefully organize and articulate their thoughts. This may discourage users from leaving comprehensive and high-quality reviews, especially when they are on the go. To address this challenge, we developed Vocalizer, a mobile application that enables users to provide reviews through voice input, with enhancements from a large language model (LLM). In a longitudinal study, we analysed user interactions with the app, focusing on AI-driven features that help refine and improve reviews. Our findings show that users frequently utilized the AI agent to add more detailed information to their reviews. We also show how interactive AI features can improve users' self-efficacy and willingness to share reviews online. Finally, we discuss the opportunities and challenges of integrating AI assistance into review-writing systems.
\end{abstract}



\begin{teaserfigure}
  \centering
  \includegraphics[width=\linewidth]{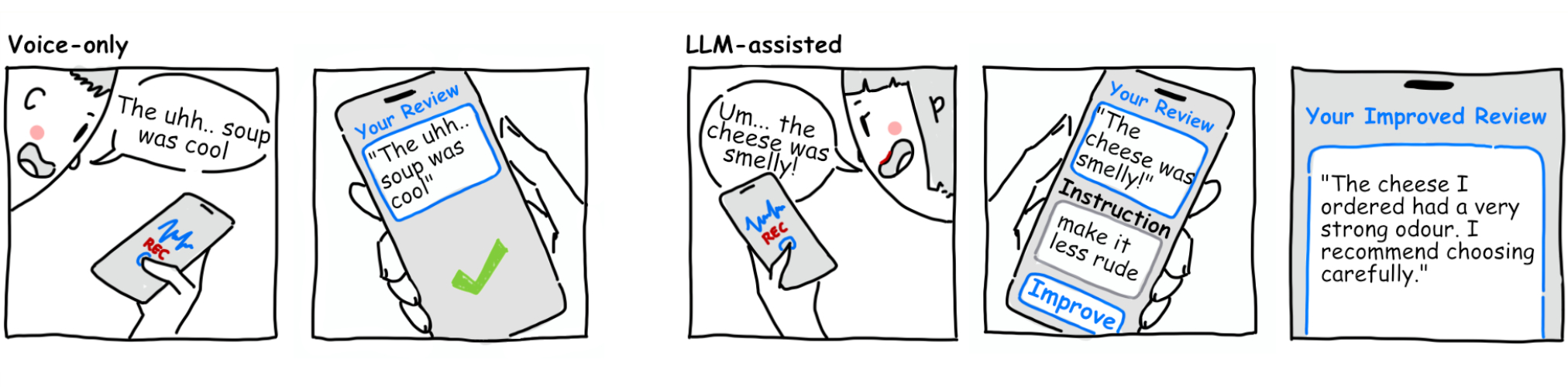}
  \caption{Usage of the voice-only (VOV) and the LLM-assisted versions (LAV) of Vocalizer. On the left: Vocalizer transcribes the spoken review. On the right: Vocalizer cleanses the review, removing unnecessary stop words and mumblings and enables the user to tweak the review with prompts or get science-based ideas for improvement.}
  \label{Fig:bannercomic}
\end{teaserfigure}

%


© Kavindu Perera | ACM 2024. This is the author’s version of the work. It is posted here for your personal use. Not for redistribution. The definitive Version of Record was published in the Proceedings of the ACM Conference on Mobile and Ubiquitous Multimedia (MUM ’24), http://dx.doi.org/10.1145/3701571.3701593.

\hfill \break

\maketitle

\section{Introduction}
Online restaurant reviews are a great resource for anyone looking for a place to eat or enjoy a cup of coffee. These user-generated evaluations not only influence dining decisions, but also shape the reputation and success of restaurants~\cite{Saddam2024, doi:10.1287/mnsc.2022.4359, LI2023104707}. Writing good reviews requires skills that improve the quality and impact of feedback. These abilities include being able to clearly articulate thoughts, developing well-informed points of view, or presentation skills~\cite{rice2007using}.

The recent commoditization of advanced voice input/output features in mobile devices (such as voice assistants and dictation features) and Large Language Models (LLMs) open up new opportunities to augment many of the skills and considerations that deal with reviews. These technologies can support human expression and improve review quality through, for instance, automatically correcting text. However, generating reviews is a different thing. For instance, a recent study has shown how AI-created reviews can be perceived as less useful, trustworthy, and authentic than human-created ones when people are aware that the content was AI-generated~\cite{AMOS2024102163}. This is a common response to AI-generated content, reflecting broader concerns about its authenticity. 


In our work, aim to strike a balance between AI-generated content and human-written content. 
To this end, we contribute \textit{Vocalizer}: an AI-powered mobile application to assist users in leaving spoken reviews conveniently on the go. We integrated a Large Language Model-powered AI agent into the review creation process, allowing users to instruct the model to edit their review that has gone through an initial automatic cleansing process. Rather than using the LLM to generate the whole review for users, we allow users to freely ramble the initial review, and then exploit AI features to enhance the review. 



Our key contributions are:

\begin{itemize}
    \item A web application that enables people to edit their restaurant reviews with an AI agent and on the go,
    \item An analysis of the strategies people prefer to employ for editing their reviews with LLM
    \item A discussion on the use of LLMs in the context of leaving good restaurant reviews, highlighting its benefits and the tensions that arise when allowing AI to modify human-made reviews.
\end{itemize}

Through a one-month field study with two Vocalizer versions (one without the interactive AI features and one with them), we found that users used the AI agent to improve the quality of their reviews by adding more information, clarifying unclear points, and removing unnecessary content. The use of AI also led to an increase in users' confidence in writing good reviews, as measured through a simple self-efficacy metric. 
Through our results, we also discuss the potential of AI to aid and improve the review writing process, making it more efficient for a wider range of users.





\section{Related Work}

\subsection{Online Reviews}
Past work has examined the role of online reviews in shaping consumer behaviour. Studies have consistently shown that online reviews influence purchasing decisions, particularly in the restaurant industry~\cite{alfira2024enhancing, haro2024influence, singh2023word}. In addition, the content and sentiment of online reviews have been shown to impact a restaurant's reputation and overall success~\cite{Saddam2024, LI2023104707}. The characteristics of the review can also significantly impact how consumers perceive the review and whether or not they trust it \cite{furner2021}. The key characteristics of a high-quality review include the following.

\begin{itemize}
    \setlength\itemsep{0pt}
    \setlength\parskip{0pt}
    \item \textbf{Detailed and informative:} High-quality reviews provide specific examples of the reviewer's experience with the product or service being reviewed. For example, instead of saying, ``This restaurant was great,'' a high-quality review might say, ``I had the steak at this restaurant and it was cooked perfectly. It was also very tender and juicy. The service was also excellent.''~\cite{mudambi2010, Liang2019,rice2007using}.
     \item \textbf{Objective and neutral:} Reviews written in a neutral and objective tone are considered quality reviews~\cite{Suzuki2020, mudambi2010,rice2007using}. They avoid using emotional language or making personal attacks. For example, instead of saying, ``I loved this product,'' a high-quality review might say, ``This product is well-made and durable. It also has a number of features that I find useful.''. 
    \item \textbf{Readable and easy to read:} High-quality reviews are well-written and easy to read~\cite{HUANG2018430}. They use proper grammar and spelling, and they are formatted in a way that is easy to scan. For example, a high-quality review might use bullet points or numbered lists to make it easier for readers to find the information they are looking for.
    \item \textbf{Recency, Length, and Quantity of reviews:} Several factors affect consumer trust in online reviews. Review recency is a key element, with recent reviews holding more credibility as they better represent the current state of a product or service~\cite{tandon2021analyzing}. Longer reviews tend to be trusted more as they offer in-depth insights and comprehensive information~\cite{mudambi2010, Liang2019, tandon2021analyzing}. Additionally, a high quantity of positive reviews contributes to increased consumer trust, signalling a larger satisfied customer base.
    \item \textbf{Sentiment of the review:} Consumers are more likely to trust reviews that have very high or low ratings because they are often seen as more helpful. This is because individuals who experience strong emotions typically put significant effort into expressing their high satisfaction or intense dissatisfaction, leading to higher-quality reviews~\cite{Liang2019, tandon2021analyzing}.
\end{itemize}   

Consumers are more likely to trust reviews that are written by other consumers who have similar demographic characteristics, such as age, gender, and location~\cite{ahn2024impact}. This is because consumers are more likely to feel that they can relate to the reviewer's experience and that the reviewer's opinion is relevant to them~\cite{ahn2024impact}. For example, a young mother is more likely to trust a review of a children's product that is written by another young mother than a review that is written by an older person who does not have children.

Studies have also investigated consumer trust and credibility in relation to online reviews~\cite{sourabh_jain_2024}, and recent findings from a non-academic survey~\cite{brightlocal2023} suggest a curious trend: consumers may in some cases even exhibit a preference for carefully crafted AI-generated review responses over those authored by humans. 

\subsection{AI-Assisted Writing}
The integration of AI, particularly Large Language Models (LLMs), into various tasks with text enables a new class of applications. Modern LLMs can now summarize complex documents, help with creative writing, or provide answers to questions~\cite{lee2024, zhao2023investigating}. Many people use commercial tools that rely on AI technologies such as LLMs to work with text. For instance, Grammarly\footnote{https://grammarly.com/} is a tool shown to help with many aspects of language editing, such as grammar, sentence structure, spelling, and punctuation~\cite{jayavalan2018effectiveness,shahriar2023potential,huang2020effectiveness}.

LLMs have uses across various domains. For example, in educational environments, LLM tools can boost students' creative abilities, enhance their writing proficiency, and increase engagement in the learning process. They help learners generate story concepts, improve grammatical precision, and collaborate with LLM systems that serve as coauthors in the creative writing process~\cite{fang2023systematic}. 

The use of LLMs extends beyond creative writing to more structured and challenging tasks such as argumentative essay writing. Studies ~\cite{li2024value} have shown that LLM can reduce cognitive strain and speed up the writing process by offering well-organized content and relevant evidence. 
This is especially useful for tasks that require a logical structure and the integration of supporting arguments, as LLM can quickly generate content that satisfies these requirements~\cite{li2024value}. The different types of writing tasks—be it creative or argumentative—highlight the various impacts of LLM assistance, notably boosting productivity in structured writing while still offering benefits in creative scenarios. 

In addition, modern smartphones have contributed to the accessibility and efficiency of LLM-assisted writing through advanced voice input technologies. Integrating voice input with AI, as seen e.g. in the \textit{Rambler} system, has been shown to improve the writing process by addressing common issues in voice dictation, such as disfluency, verbosity, and incoherence~\cite{Lin2024}. Users of \textit{Rambler} reported that the system facilitated the development and refinement of spoken content, supported iterative revisions, and enhanced user control during editing. This synthesis of spontaneous speech and organized writing exemplifies how LLMs can bridge the gap between natural language expression and formalized text production.

Although advances have been made, the development of AI-generated text presents challenges related to factual accuracy, bias, and the ethical consequences of creating synthetic content. The growing sophistication of AI technologies in producing text that resembles human writing prompts important questions about the authenticity and dependability of such content~\cite{marco2024pronvspromptlarge}. This issue is particularly noticeable in areas like online reviews, where it is becoming harder to distinguish between content created by humans and AI. Recent research indicates that human reviewers often find it difficult to separate AI-generated restaurant reviews from those written by humans~\cite{gambetti2023combat}. This raises ethical issues in situations where the authenticity and humanness of content are important.


Concerns about the potential misuse of AI in generating misleading content are increasing. Research by Bartoli et al.~\cite{bartoli2016best} showed that it is feasible to produce hundreds of fake reviews designed for specific ratings and restaurant categories, successfully deceiving users. Their experimental study with 39 human participants showed that about 30\% of the AI-generated reviews were considered useful by human users, and roughly 25\% of the time. AI-generated reviews contradicted the sentiment of genuine reviews, resulting in scenarios where AI-generated content was considered more credible than authentic reviews.

The implications of these results are significant, especially as sophisticated language models get better in producing convincing and fake reviews. This development challenges the trustworthiness of user-created content on social media sites, highlighting the necessity for strong measures to identify and reduce the effect of AI-generated fake reviews~\cite{gambetti2024aigen}. In our study, we seek to investigate how to harness AI to edit, not create, reviews.

\section{Methodology}
We developed the mobile application, named \textit{Vocalizer}, as a web application which was then evaluated through a deployment study conducted on a university campus with several different restaurants.

\subsection{Vocalizer: A Voice-Based AI Review Application}  
\begin{figure}[h]
  \centering
  \includegraphics[width=0.9\linewidth]{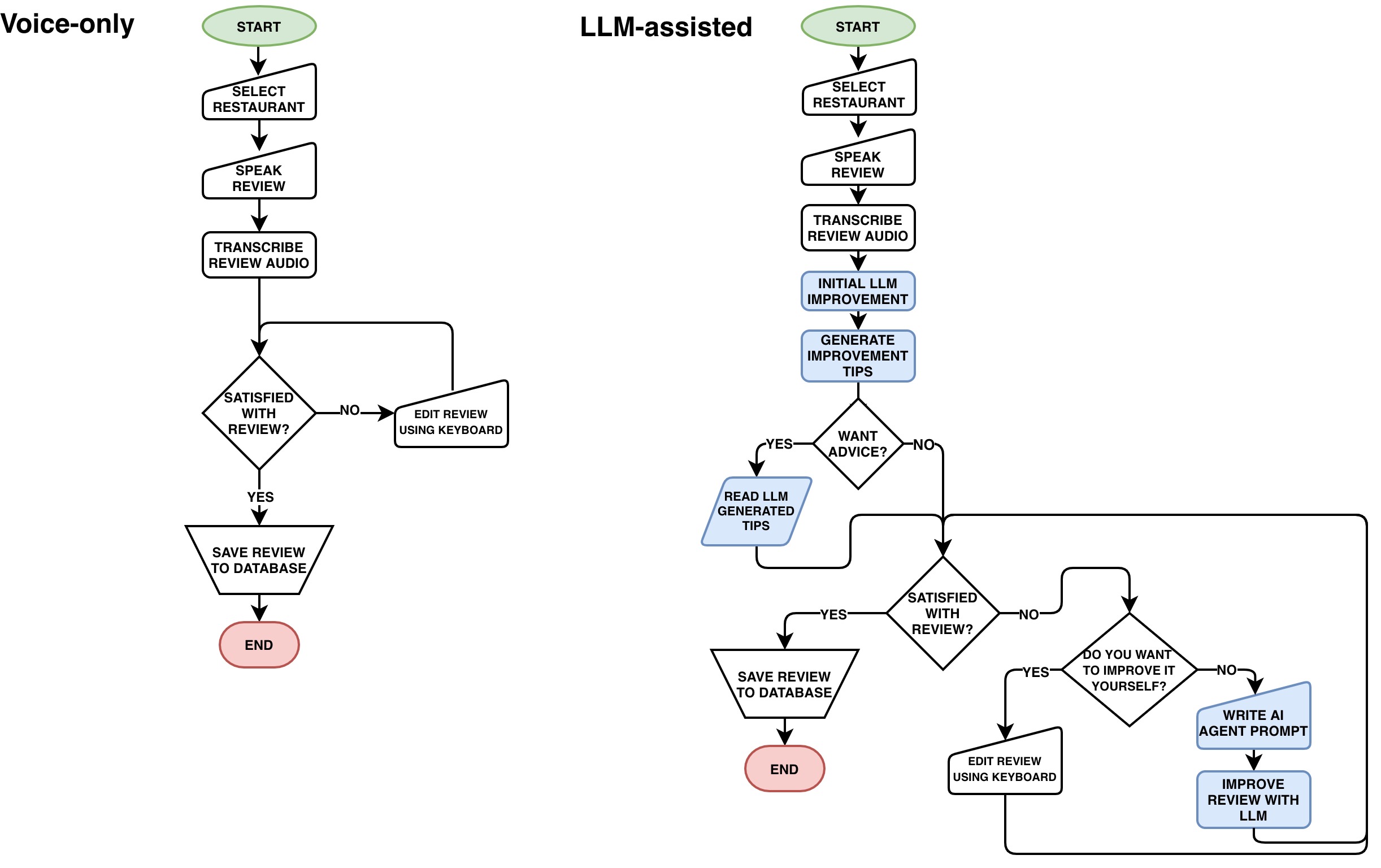}
  \caption{Flowcharts showing the operation of both versions of Vocalizer. On the left, the voice-only version, and on the right, the LLM-assisted version, with the additional LLM-powered processes and user-LLM interactions coloured blue. As shown, the LLM-assisted version offers advice the user can read or ignore, and they can prompt the AI agent to enhance the review if they do not want to edit it themselves.}
  \label{fig:flowchart}
\end{figure}

For the purpose of the study, the application was designed with two versions that acted as the study conditions: Voice-Only Version (VOV) and LLM-Assisted Version (LAV). The VOV uses only voice input, while the LAV incorporates voice input alongside a LLM. The overall operation of Vocalizer is outlined in \autoref{fig:flowchart}. We used HTML for front-end development and Express\footnote{https://expressjs.com/} and Node.js\footnote{https://nodejs.org/en} for server-side functionality.

\subsubsection{Voice-only Version}
In the VOV, participants were asked to record their restaurant reviews via voice input after visiting a restaurant. The reviews were then immediately transcribed using the OpenAI Whisper model\footnote{https://openai.com/index/whisper/} and presented to users for review and editing (see \autoref{Fig:ai-agents}-C), allowing them to correct any transcription errors and refine the content as needed. 

A dynamic waveform animation was added as visual feedback during the recording process (see \autoref{Fig:ai-agents}-B). These visual elements can improve user motivation and task participation, potentially leading to more detailed and thorough reviews~\cite{Fernandes2018}. We set a maximum of 5 minutes of recording time due to technical considerations, but the subtle countdown timer was shown after 30 seconds to create a relaxed and unrestricted recording environment while still ensuring sufficient data collection.
Before submitting the review, users were asked for quick feedback using a slider input (scale 1--7, from \textit{not at all willing} to \textit{extremely willing}) about their willingness to share the review online.


\subsubsection{LLM-Assisted Version}
The LAV builds on the voice-only version by including various language improvement functionalities, as detailed below, through the OpenAI GPT-4\footnote{https://openai.com/index/gpt-4/} model. A fundamental design principle in creating the LAV for the Vocalizer application was highlighting user autonomy and control. 

\begin{figure}[h!]
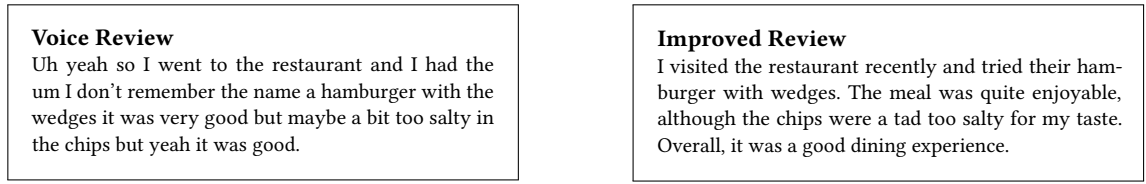

    \centering
    \begin{minipage}[t]{0.45\textwidth}
        \begin{framed}
            \textbf{Voice Review} \\
            \small{Uh yeah so I went to the restaurant and I had the um I don't remember the name a hamburger with the wedges it was very good but maybe a bit too salty in the chips but yeah it was good.}
        \end{framed}
    \end{minipage}
    \hfill
    \begin{minipage}[t]{0.45\textwidth}
        \begin{framed}
            \textbf{Improved Review} \\
            \small{I visited the restaurant recently and tried their hamburger with wedges. The meal was quite enjoyable, although the chips were a tad too salty for my taste. Overall, it was a good dining experience.}
        \end{framed}
    \end{minipage}
    \caption{Comparison of an example original review and its AI-enhanced revision.}
    \label{Fig: voice-and-improved}
\end{figure}

\textbf{Initial Review Improvement:} LAV refines the spoken review automatically by rephrasing sentences for improved clarity, correcting grammatical errors, removing unrelated content to ensure coherence, and eliminating filler words like `um' and `ah' while retaining the original content and tone (see Figure~\ref{Fig: voice-and-improved}-B).
As spoken text tends to be "messy" by nature, we sought to avoid the user having to rely on first using the AI agent to remove the filler words before considering more substantial edits to the review.
AI agent can also identify if the user goes off-topic or talks about irrelevant details as part of the review, and remove these sections. 
The LAV version converts raw transcripts into polished text that is shown on the screen accompanied by a voice transcription. Users can compare the original voice transcription with the refined review text.

\begin{figure}[h]
  \centering
  \includegraphics[width=0.9\linewidth]{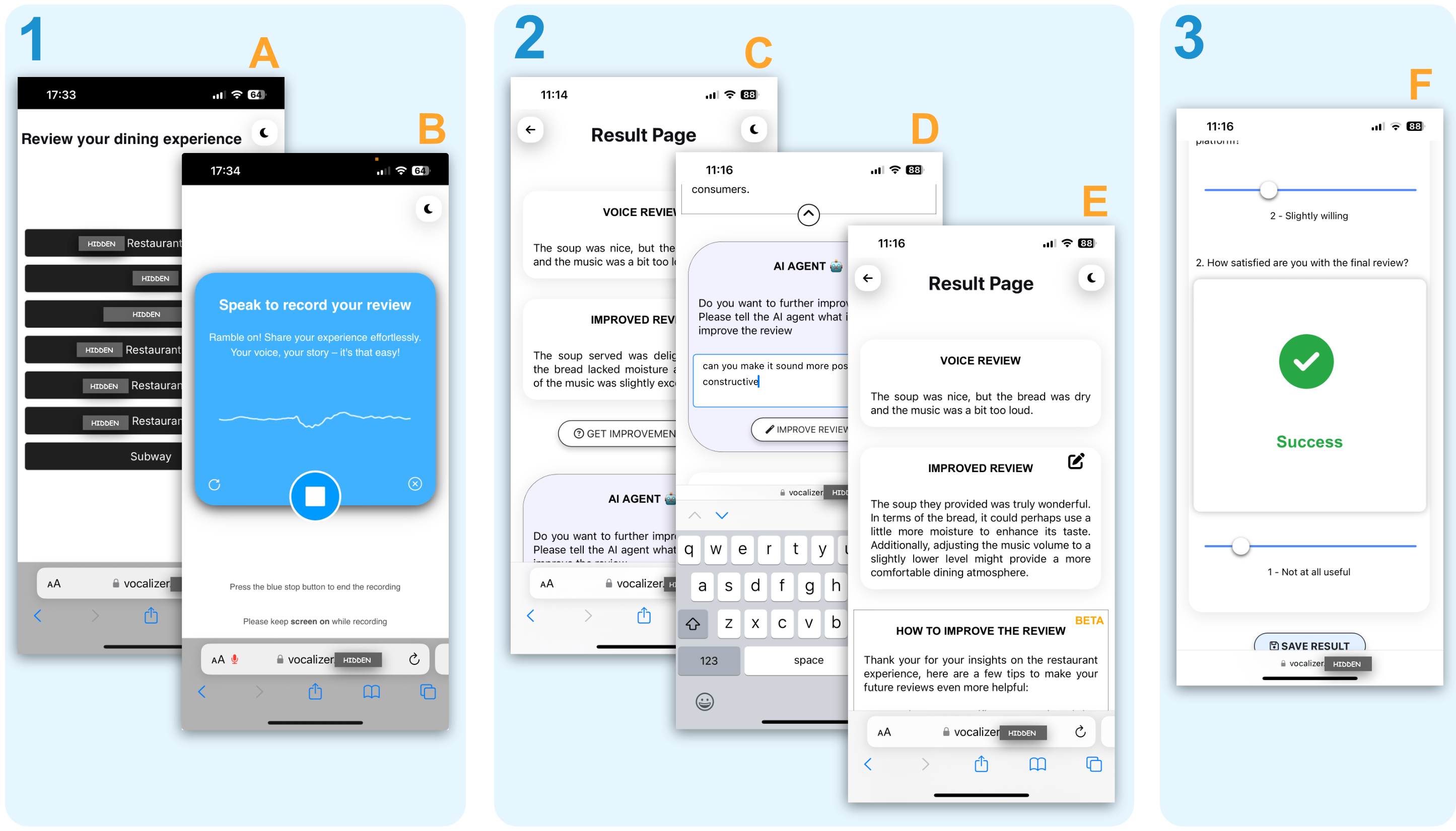}
  \caption{Core functionalities of the LLM-Assisted Version of Vocalizer presented chronologically from left to right. 1 - Selecting restaurant (A) and recording review (B), 2 - reading transcription and improved version (C), asking AI agent to improve review with a text prompt (D) then reading the newer iteration of the improved review with personalised, AI-generated improvement ideas visible below (E), 3 - entering feedback on sliders and submitting restaurant review (F)
  }
  \label{Fig:ai-agents}
\end{figure}

\textbf{AI Agent:} A conversational AI assistant (see Figure~\ref{Fig:ai-agents}-C, D) provides the ability to further edit the improved review through natural language commands typed by the user on the device's keyboard. This functionality is designed to study how users leverage AI to improve and customize their reviews.
As the user instructs the AI agent on how to modify the review, it employs the GPT-4 model to make the desired adjustments (see \autoref{Fig:bannercomic}, panels 4 and 5). The user is allowed to repeat this step and iteratively improve their review until they are satisfied with the result.

\textbf{Review Improvement Tips:} In order to improve the quality of user-generated reviews, our application offers enhancement suggestions based on the features of high-quality reviews that had been identified in related research. We provide the findings to the LLM through prompts, as detailed in the listing \autoref{lst:suggestion_prompts}. When users request assistance with review improvement at the tap of a button, the AI agent shows suggestions specific to their review as shown in Figure \ref{Fig:ai-agents}-E. For example, the brief review \textit{`the soup was very tasty'} may receive the following advice: \textit{`Your review is short and sweet, and it's great that you mentioned the positive aspects of the dish. However, to enhance its quality, consider providing more details. You could specify what type of soup it was, and what made its flavor so delightful. Talk about the ingredients or any unique flavor it had. You could also elaborate on the restaurant's service or atmosphere to make the review more comprehensive. Remember that more explanatory and in-depth reviews tend to be more useful for readers'}. 

Furthermore, users can restart the review process and go back to the beginning if they find the agent's revisions of their review unsatisfactory. All interactions are logged within the application for later examination. 

Again, before submitting each review, participants were required to leave fast in-app feedback, as illustrated in Figure~\ref{Fig:ai-agents}-F. This time, in addition to the item about willingness to share, the users were also asked about their overall satisfaction with the review, if the provided ideas were helpful (in case they used the suggestions module) and if the AI agent was useful in improving the review. The voice-only version contained one feedback question (share willingness), whereas the LLM-Assisted Version featured up to four questions (share willingness, review satisfaction, improvement idea helpfulness, AI agent usefulness) to assess the various AI components according to the levels of feature interaction. All feedback collected was stored in an online database for later evaluation.

\subsection{Prompt Design}
The efficacy of AI models depends on the specificity and clarity of the prompts~\cite{Talarico2023}. This study used a systematic approach to prompt engineering, with the aim of optimizing AI performance by carefully considering prompt structure and content. 
The variable \texttt{generatedText} refers to the review that has already undergone initial improvement by the LLM, eliminating redundant information to improve clarity. The variable \texttt{refineInstructions} refers to the instructions given by the participants on how to refine this review. The GPT model receives the texts represented by these two identifiers along with their brief descriptions before receiving the prompt to process them. 

Although Vocalizer appears to be one cohesive AI system to the user, it is made up of three separate but complementing LLM-powered features that operate independently. In the following, we describe the differences in how the ChatGPT model is utilised for these three features. Table ~\ref{tab:ai_comparison} presents the inputs, system prompts, and temperature settings for each AI feature. These three specify which textual data a certain LLM step in the system works on, what system prompt it receives (an essential part in making models like ChatGPT behave in specified ways) and what temperature the model runs on (how creative the model is allowed to be at the expense of factual accuracy) respectively.

\begin{table}[h!]
\centering
\resizebox{\linewidth}{!}{%
\begin{tabular}{p{1.5cm}p{2.8cm}p{5cm}p{5cm}}
\toprule
\textbf{Parameter} & \textbf{Initial Review Improvement} & \textbf{AI Agent} & \textbf{Review Improvement Tips} \\ \midrule
\textbf{Input} & The transcribed voice text \& Restaurant name & User-given text in the instructions box of AI agent and Improved review text & Improved review text \\ 
\textbf{System Prompt} & You are a restaurant review improve GPT & I want you to act as a text refinement agent. You should read the \texttt{generatedText} review and the instructions given by the user. Then refine the review of how the user asked in the user \texttt{refineInstructions}. Do not use any fancy English words, use a basic English level that everyone can understand. & I want you to act as an adviser. You need to read \texttt{generatedText} restaurant reviews and give some tips and tricks to improve their review quality. You should only focus on facts that are confirmed by scientific research but do not add any reference to the source. I added all the scientific findings on how a high-quality review looks in the prompts. \\ 
\textbf{Temperature} & 0.2 & 0.8 & 1 \\ \bottomrule
\end{tabular}%
}
\caption{Comparison of Parameters across different AI features}
\label{tab:ai_comparison}
\end{table}

These three features rely on further instructions that ground the LLM output in research and prevent certain erroneous behaviors to ensure consistent, reliable and correct output from the system.

First, in order to minimize the risk of hallucination and align with the voice review in terms of facts and overall sentiment, specific restrictions referenced in \autoref{sec:appendix_a} were applied to the initial LLM improvement in the LAV. These instructions enabled keeping the original tone and content unchanged, while excluding filler words. 
Second, the list shown in \autoref{sec:appendix_b} provides guidelines for an AI agent tasked with enhancing user-created reviews. 
Finally, an LLM-powered module for review improvement tips was added to give users advice specific to their own reviews. This feature offers suggestions grounded in research on review writing. The list in \autoref{sec:appendix_c} shows how we included the most recent research insights into the bot to create customized improvement tips.

\subsection{Study Design}
\begin{figure}[h]
  \centering
  \includegraphics[width=0.8\linewidth]{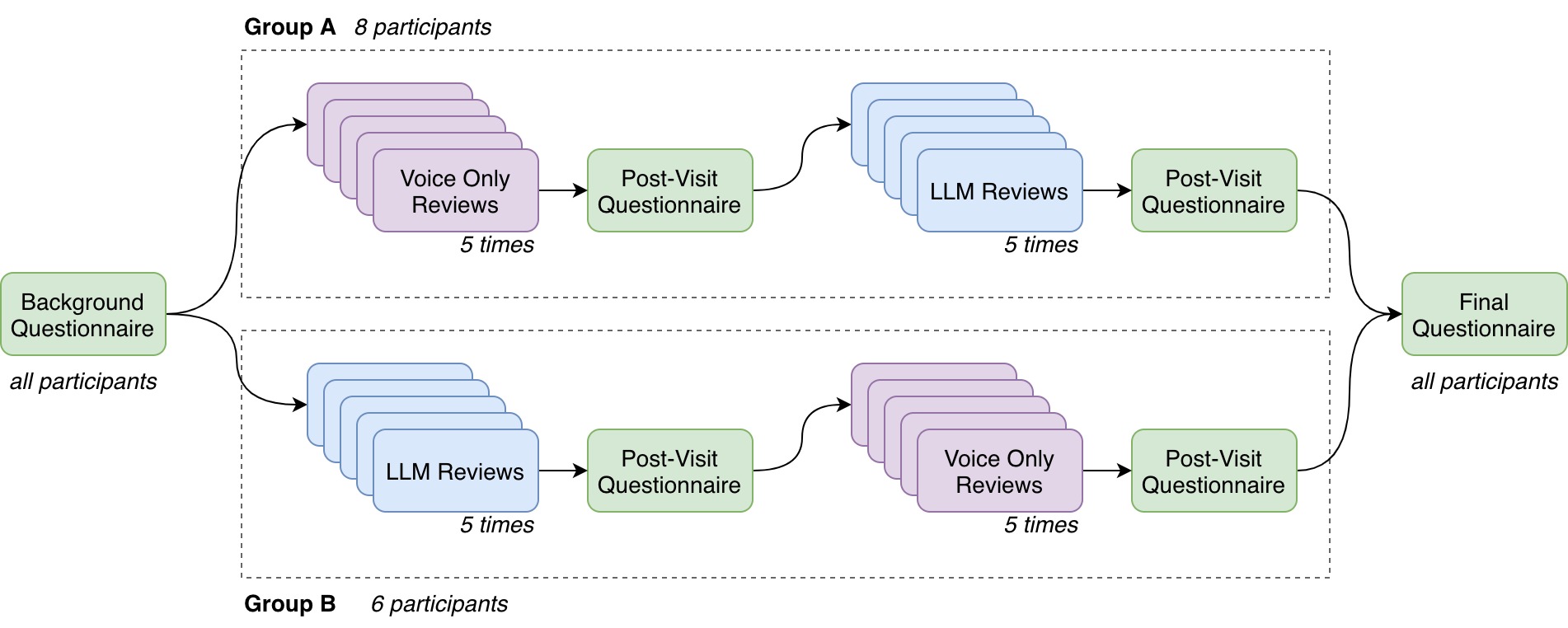}
  \caption{Study design diagram showing the order of tasks and questionnaires for each participant.}
  \label{Fig:study-design}
\end{figure}

We used a mixed method research design to explore the use of AI to improve reviews. We captured data directly in the application and using a set of three questionnaires that were administered using Google Forms. The overall study design is shown in \autoref{Fig:study-design} and in the following, we detail the exact data collection methodology.

\subsubsection{Participants}
We recruited 14 participants, 7 women and 7 men, with an average age of 28.21 years (SD = 5.32).
Participants were recruited through the university's research participation system and on-campus posters. 
This online participation system allows researchers to advertise a study and allows potential participants to register for projects that align with their interests.
Each participant was promised approximately \$20 worth of university merchandise as a reward for their time.

\subsubsection{Protocol}
We used a counterbalanced within-subjects design, as shown in \autoref{Fig:study-design}. Participants were randomly assigned to begin the study period with either the VOV or the LAV condition. The lead author sent each participant a welcome email that contained instructions on how to start, how to install the application, and what their tasks were. The first task was to complete at least five restaurant reviews from any of the campus restaurants. After completing the first five reviews and the corresponding post-visit questionnaire, they continued the study with the alternate version of the application. Participants were asked to email the lead author after finishing the reviews with either version of the application. The lead author kept a record of participants' progress to confirm that the participants proceeded according to protocol. The participants did not interact with Vocalizer outside the context of this study, and all participants completed the study period within three weeks. 

\subsubsection{Questionnaires}
The background questionnaire included items on demographic information, prior experience with restaurant reviews and review platforms, and attitude toward restaurant reviews. The questionnaire was presented prior to the installation of the application, in accordance with the guidelines provided in the welcome email.

After each review, we collected rapid in-app feedback, as follows: 
\begin{itemize}
    \item Q1: How willing would you be to share the final review online in a public restaurant review platform? (VOV \& LAV)
    \item Q2: How satisfied are you with the final review? (LAV)
    \item Q3: Were the provided improvement ideas helpful (LAV in case the ``Get Improvement Ideas'' functionality was used)
    \item Q4: How useful was the AI agent in improving your review? (LAV)
\end{itemize}

After completing five reviews with each version of the application, participants received a post-visit questionnaire. This questionnaire incorporated the short form of the user experience questionnaire based on a 7-point scale (UEQ-S)~\cite{schrepp2017design} to assess user experience. In addition, we were interested in people's self-efficacy in leaving reviews using the application. This metric originates from Bandura's theory of social learning~\cite{bandura2006guide}, which posits an individual's belief in his ability to perform a task successfully. To that end, we posed the question: \textit{Please rate how certain you are that you can use the application to give a good (useful and enjoyable) restaurant review}. The responses were recorded on a likert scale of 0 to 10 to evaluate the self-efficacy of the participants after each version of the application.

A final questionnaire was administed in the end of the study. The questionnaire first explored the participants \textit{overall} perception of their self-efficacy in writing reviews with or without Vocalizer, now that they had used it thoroughly. Here, we used the same scale as in the post-visit questionnaires (0-10) for easier comparison. The questionnaire also contained open-ended items on the specific ways Vocalizer can help in review writing and on explaining which of the versions they preferred and why. We also allowed the participants to leave free comments, and in the end asked a more broad question about the threats and opportunities they can think of in using AI in the review process.  

\section{Results}

\subsection{Overview of the Review Experience of Participants}
Participants reported eating on campus frequently, with an average of almost four visits per week (most restaurants are closed on weekends except for one fast-food spot), showing a significant engagement with the campus food scene. 
Participants rated their review experience on average 3.43 (SD~=~1.91) on a scale from 1--7, indicating a range of levels of expertise. Google Reviews was the most used platform (\textit{N}~=~10), but the participants also mentioned using TripAdvisor, restaurant websites, Facebook, Reddit, Jodel, and Yelp to collect and share experiences. Results about favourite features on review platforms showed that traditional and lower-effort methods may be desirable. Four of the 14 participants mentioned that multimedia reviews are particularly important for them, and at least five thought that short, easy review formats such as stars are appreciated.

\subsection{Vocalizer Usage Analysis}

In the Voice-Only Version, we collected 75 reviews. 56 submitted the original transcript, while 19 reviews were manually edited by users, with an average of 2.11 edits per review among the latter. 



In the LLM-Assisted Version, we collected a total of 82 reviews. These reviews have all been improved using LLM after transcription of the voice input. Examples of how users' transcribed reviews were enhanced at the initial LLM improvement step are shown in \ref{sec:appendix_d}. Out of these, users asked the AI agent to further modify 42 reviews, while the rest were submitted without the use of the AI agent. Users requested recommendations to improve their reviews 20 times by pressing the button 'Get improvement ideas' shown in Figure \ref{Fig:ai-agents}-E. Examples of reviews and the improved advice they received are shown in \autoref{sec:appendix_e}.
Upon receiving suggestions, 14 reviews were further refined: 11 were refined with the help of the AI agent, and 3 with manual revisions (writing with the keyboard).
Users gave feedback on the usefulness of the AI agent's review improvement suggestions using a simple thumbs-up mechanism.

\begin{figure}[h]
  \centering
  \includegraphics[width=1\linewidth]{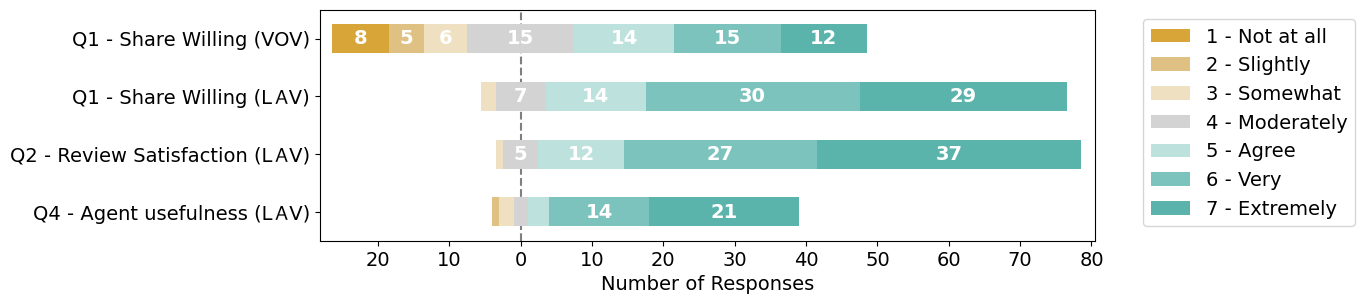}
  \caption{Feedback response distribution for evaluating the VOV and LAV. The figure illustrates user perceptions regarding the usefulness, overall satisfaction, and willingness to share reviews generated by application versions.}
  \label{Fig: willingness-vs-share}
\end{figure}

Figure \ref{Fig: willingness-vs-share} compares the VOV and LAV feedback data.

\subsubsection{Q1: Willingness to share [VOV \& LAV]} 

A paired-sample t-test was performed to compare the willingness of the participants to share reviews produced by VOV and LAV. The results indicated a significant difference between the two conditions. Participants reported a higher willingness to share LAV reviews (M = 5.93, SD = 1.04) compared to VOV reviews (M = 4.53, SD = 1.85), t(14) = -2.39, p < 0.05. This suggests that the LAV was more effective in producing content that met user expectations for public sharing.

\subsubsection{Q2: Review satisfaction (LAV)} 

Participants showed a high level of satisfaction with the reviews provided by the LLM-Assisted Version (N = 82). The average satisfaction score was 6.15 out of 7, with a standard deviation of 0.96, reflecting a positive reception of the quality of the review. 

\subsubsection{Q3: Usefulness of improvement ideas by the AI agent}
All of the 20 improvement suggestions were considered helpful, as indicated by thumbs up or down.

\subsubsection{Q4: Usefulness of the AI agent (LAV)}

The mean rating for agent usefulness was 6.09 (SD = 1.24, N = 42), indicating a generally positive perception.
Although a majority of users expressed high levels of satisfaction (categories ``Very'' and ``Extremely'' useful accounted for 83.33\% of responses), a smaller subset provided more moderate ratings. This indicates that although the AI agent was beneficial to most users, there could be particular situations or individual traits that affected its perceived usefulness.

\subsection{Strategies for Revising Prompts for a Better Review}
\label{sec: qualitative_prompts}

We adapted Braun and Clarke's thematic analysis~\cite{braunthematic} to fit our smaller team. The lead author and a coauthor first reviewed the prompts that the participants used to modify their reviews. Then, we collaboratively created the initial codes (Modification Types) using Google Sheets as the shared platform. In the second step, the codes were reworded when necessary and merged into broader themes (Categories). We noticed a dominant focus on \textit{content modification}, with users seeking to add, remove, or rephrase information. In addition, users frequently requested \textit{sentiment adjustment}, shifting between positive, negative, or neutral tones. \textit{Style control}, such as improving readability or conciseness, was also observed. The types of modifications and their frequencies can be seen in Table~\ref{tab: user-promts}.

\begin{table}[ht]
  \caption{Distribution of User Prompts}
  \label{tab: user-promts}
  \centering
  \footnotesize
  \begin{tabular}{ccl}
    \toprule
    \textbf{Category} & \textbf{Modification Type} & \textbf{Freq.} \\
    \midrule
    \multirow{5}{*}{Content Modification} & Omit Information & 8 \\
           & Add Details & 30 \\
           & Correct Information & 5 \\
           & Focus Adjustment & 2 \\
           & Clarify Ambiguity & 10 \\
           
    \midrule
    \multirow{6}{*}{Sentiment Adjustment} & Formality Increase & 3 \\
                                          & Formality Decrease & 6 \\
                                          & Positivity Increase & 2 \\
                                          & Positivity Decrease & 1 \\
                                          & Increase Excitement & 4 \\
                                          & Decrease Excitement & 2 \\                                        
    \midrule
    \multirow{3}{*}{Style Control} & Readability Improvement & 5 \\
    & Length Adjustment & 1 \\
    & Platform Adaptation & 2 \\
  \bottomrule
\end{tabular}
\end{table}



Table~\ref{tab: Prompts} provides examples from different review categories, displaying both the initial user reviews and their cleaned-up versions generated by the language model according to the user instructions. The `Original Review' column contains the initial review, whereas the `Cleanup Review' column displays the revised content produced by the LLM based on the given instructions. The `Instructions' column shows what the users wanted to edit (and the modification type assigned in our own analysis).




\subsection{Impact of Reviewer Experience on Prompt Length}
\label{sec: experience-length}
We observed a positive correlation between the prior experience of users in leaving reviews and the length of the prompts they provided to the AI assistant. More experienced users tended to use longer prompts compared to less experienced users (Spearman correlation = 0.94, p < 0.01). Although further research into this is warranted, we believe that more experienced users simply know better what they want and how to write good reviews. 

\begin{figure}[h]
  \centering
  \includegraphics[width=0.75\linewidth]{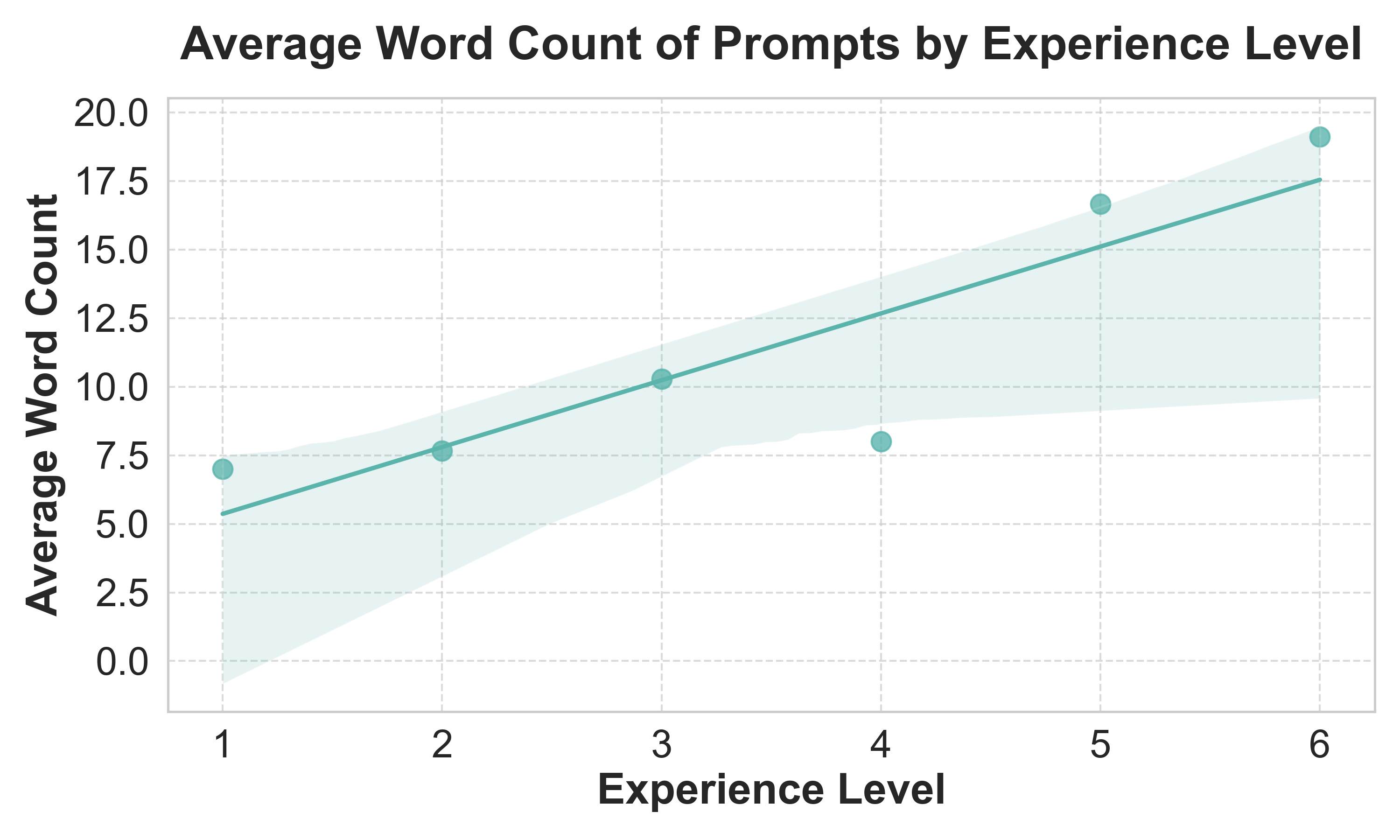}
  \caption{
Scatter plot illustrating the relationship between participants' self-reported experience levels in reviewing restaurants and the average word count of the prompts they provided. The regression line shows the trend across different experience levels, highlighting the correlation between experience and the length of instructions.}
  \label{Fig: word-count}
\end{figure}

\subsection{User Experience Analysis}

\begin{figure}[h]
  \centering
  \includegraphics[width=1\linewidth]{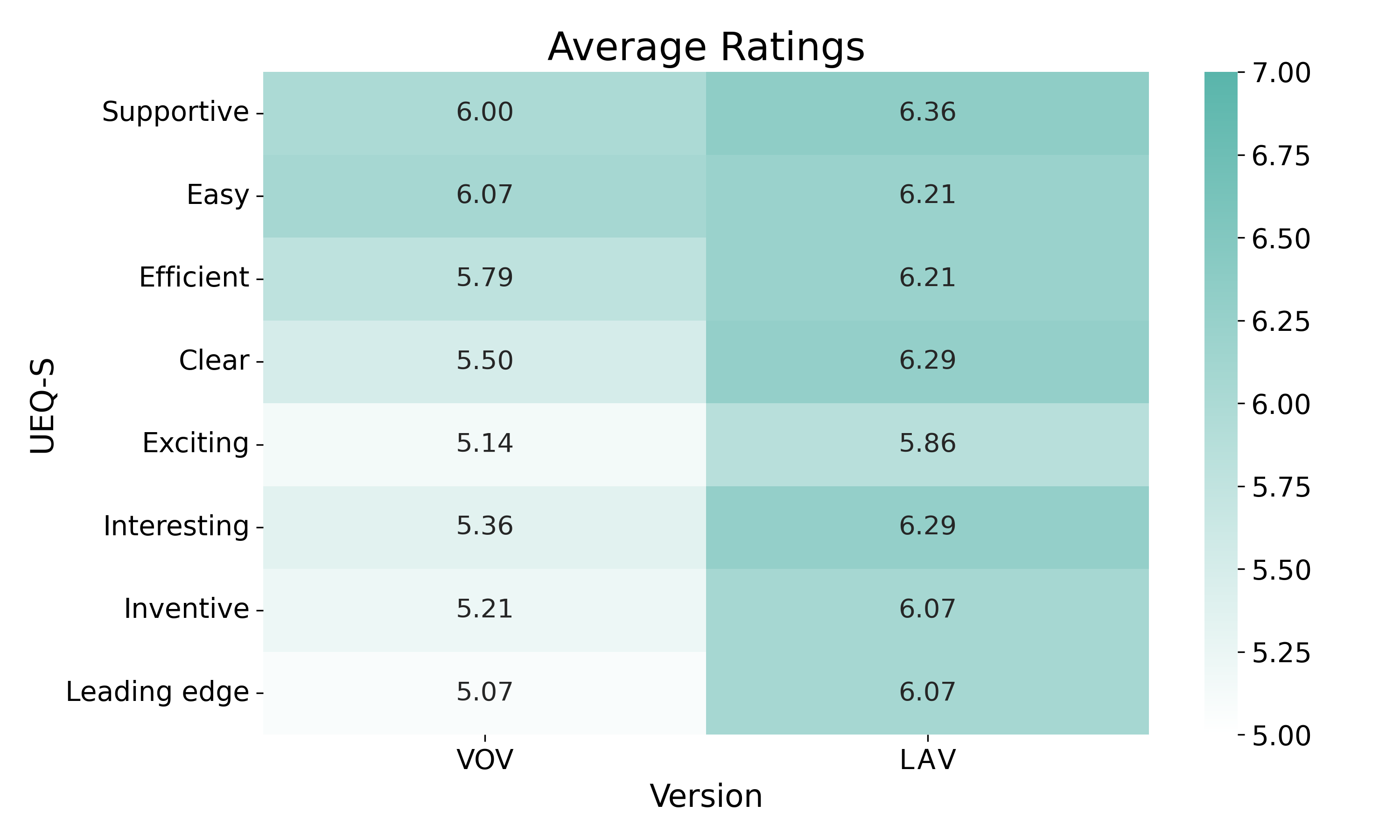}
  \caption{Results of the User Experience Questionnaire Scores for VOV and LAV. The heatmap displays the average ratings across various pragmatic and hedonic quality dimensions for both the LLM and audio-based review tools.}
  \label{Fig: ueq-ratings}
\end{figure}

Our post-visit questionnaires, using a 7-point Likert scale (1 = strongly disagree, 7 = strongly agree) based on the UEQ-S questionnaire, provided information on user experience with the VOV and LAV interfaces on our restaurant review platform (N = 28). The LAV seemed to outperform the VOV in all hedonic and pragmatic quality dimensions, as illustrated in Figure \ref{Fig: ueq-ratings}. The LAV achieved an average UEQ-S score of 6.17, compared to 5.52 for the VOV. The pragmatic score was 6.27 for LAV and 5.84 for VOV, while the hedonic score was 6.07 for LAV versus 5.20 for the VOV. However, none of these differences were statistically significant, as determined by Mann-Whitney U tests with Bonferroni corrections applied (adjusted significance threshold = 0.00625).


\subsection{Self-Efficacy for Writing Reviews}
\label{sec: self-efficiency}

\begin{figure}[h]
    \centering
        \centering
        \includegraphics[width=0.40\textwidth]{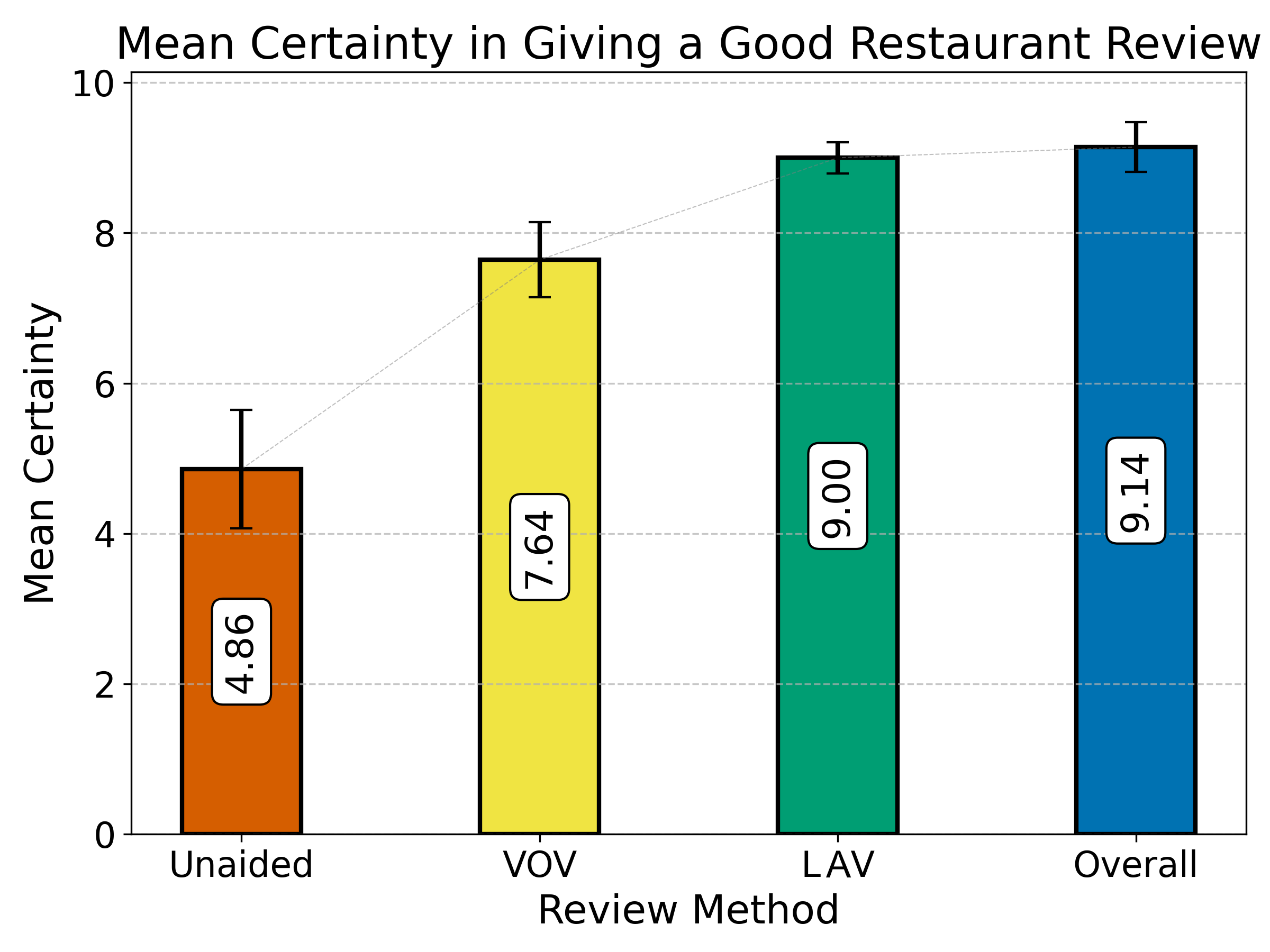}
         \caption{Self-efficacy scores for writing engaging and informative reviews before, during, and after utilizing the AI tool. Ratings for VOV and LAV were gathered from the post-visit questionnaire to capture immediate impressions after trying each version. A final questionnaire assessed user self-efficacy with and without the aid of a tool like Vocalizer. Ratings without the tool are referred to as Unaided, while those with the tool are referred to as Overall.}
        \label{Fig: mean-certainty}
\end{figure}

We evaluated self-efficacy on a scale from 0 to 10 in four aspects. First, we asked participants in the post-visit questionnaires to understand how each of the two conditions (VOV and LAV) help people. Second, in the final questionnaire, we also asked participants about their self-efficacy when they were not using any of these novel tools. Finally, beyond the two cases compared in this study, the final questionnaire was used to understand what people think about the overall concept of voice-input review creation (i.e. the concept of spoken reviews via a mobile application, with or without LLM assistance, looking beyond the implementation used in our study). Thus, we measured their self-efficacy in four aspects: \textit{Unaided} review creation, Vocalizer \textit{VOV}, Vocalizer \textit{LAV}, and \textit{Overall}. An overview of the development of people's self-efficacy is depicted in Figure~\ref{Fig: mean-certainty}

As illustrated in Figure \ref{Fig: mean-certainty}, participants reported an average self-efficacy score of 4.86 out of 10 without any support (labeled as ``Unaided'' in Figure \ref{Fig: mean-certainty}). The self-efficacy score reported by participants after using the VOV was 8.25, whereas it was 9.00 after using the LAV. An overall score (i.e. considering the whole concept, in the final questionnaire) was 9.14. To this end, we conducted a Friedman test, a non-parametric test for repeated measures to evaluate the differences. We observe a significant difference in self-efficacy scores across conditions, $\chi^2(3) = 20.15, p < 0.001$. To explore which conditions differed significantly, we performed pairwise post-hoc comparisons using the Wilcoxon signed-rank test with a Bonferroni correction to account for multiple comparisons. We observe significant differences between the LAV condition and the Unaided condition (i.e. when people estimated their ability without any tools, in the final questionnaire) ($p < 0.001$) as well as between the Unaided and Overall (i.e. with a tool like ours) conditions ($p < 0.001$). This suggests that the LAV increased users' self-efficacy in their review writing abilities.

\subsection{Qualitative Analysis of Post-Study Questionnaire}
Following the same method described in Section \ref{sec: qualitative_prompts}, we analysed the responses in the Final Questionnaire. 
Most participants shared positive experiences regarding the application.

\subsubsection{How did Vocalizer help, or did not help, leave reviews}
The first part of the analysis looks at easing the process of leaving reviews.\\
\textbf{Ease and efficiency.} First, almost all participants highlighted the effortlessness of verbal reviews enabled by Vocalizer. Users consistently highlighted that the process was easy and fast compared to traditional written reviews. As P17 noted, \textit{``The speech-to-text feature saves time and is way more low-effort compared to written reviews. You can just explain the experience out loud as you would to a friend, and then you can just fix it up with the AI features. Genius.''} This preference for voice review indicates a trend toward minimizing the cognitive and physical effort involved in the review process \cite{Miriam2020}. Users appreciated the ability to articulate their experiences verbally without the need to focus on typing or grammatical accuracy, making the review process more accessible and less time-consuming. 

\textbf{Transform spontaneous thoughts.} The LLM-Assisted Version was also recognized for its ability to turn ideas and short comments into well-structured, informative reviews. Nine participants mentioned that the AI applications made the review leaving process more efficient by improving the overall quality of their submissions. P11 remarked, \textit{``It is easier to leave a review as you can say the review in an unstructured way and the AI makes it structured.''}. P5 repeated this sentiment, stating, \textit{``It is satisfying to see how this application turns your own random thoughts into a well-written, informative, and engaging piece of text.''} The ability of the AI to streamline the review process while enhancing the clarity and coherence of the content was a key factor in the positive reception of the tool.

\textbf{Motivational Impact.} A few participants referred to how the tool encouraged them to leave more reviews. P14 expressed, \textit{``When I use this app, it motivates me to do a review. I don't need to think about grammar mistakes, and the AI-given review is very good.''} This suggests that the AI's ability to simplify the review process and alleviate concerns about writing quality can potentially motivate people.

\subsubsection{Which version did users prefer, and why}
We asked participants to tell us how they felt about each of the two versions of Vocalizer they tried, and to explain which one they preferred and why. 

\textbf{Easier articulation.}
Every participant expressed a preference for the LAV. They highlighted several advantages of the LAV, including its ability to organizing and articulating user thoughts more effectively than the audio-only version. P2 exemplified this with a comments \textit{``So it gathers your thoughts properly and has the function of changing the phrases as you want.''}, and another participant stated \textit{``I prefer to use the one with an AI agent. The AI agent will help me for the grammatical error, and clearly make the sentence which I really want to say.''}

\textbf{Dialogue.}
The other clear reason for the participants' preference was the capability to have a two-way conversation about the review with the AI. Here, P15 said that \textit{``It was fun to experiment with different change requests and see how the ai interpreted it.''}


\subsubsection{Foreseeable opportunities and threats in using AI for review process} 
\textbf{Authenticity.} First, several participants mentioned their concerns about the AI-enhanced reviews' ability to convey the author's true feelings, or in other words, their authenticity. They thought that while the LLM makes an objectively better-written review, it loses some of its meaning and nuances may be lost in the process: \textit{''... there might be a difference in the intended feedback and AI-version of it''} (P11), and \textit{``if AI corrects a lot of the sentences, that will hide the people`s honest feelings''} (P12).

\textbf{Societal Impact.} Second, although these may not be specific to user-generated reviews and review platforms, participants pointed out several negative implications on society. 
Two participants explained that AI taking over small human tasks like this will harm individuals' thinking abilities:  \textit{``our imagination and creativity will not improve''} (P1) and \textit{``it suppresses our critical thinking ability. Coming to restaurants, yeah it might be easy to give review but proof reading it will be better.''} (P4)
One participant, P9 highlighted potential malicious use of the technology, as \textit{``people find it too easy to write restaurant reviews that they would say malicious things without thinking, making it easier to defame restaurants without a justified cause.''}.

\textbf{User Experience}
On the positive side, participants thought the use of AI has the potential to enhance user experience in leaving reviews. For instance, comments like “\textit{The whole experience was pleasant and fun}” (P8) and “\textit{It was really fun! I enjoyed rambling on my review}” (P12) reflect how using AI to assist in the task added an element of enjoyment to the experience.

\textbf{Bridging Language Barriers.}
The potential of AI in enhancing communication was noted by users who could empathize with the struggle with language barriers or writing skills. Participants noted that AI can not only help in writing reviews, but as a byproduct teach people language skills. This was highlighted by comments such as “\textit{Opportunities: people with broken English can create well written reviews}” (P6) and “\textit{ I think I would like to have the AI agent as my writing tutor, because it can help me generate ideas and improve my paragraphs by suggesting which areas to elaborate}” (P9). It is thus very probable that AI can advance accessibility and inclusion by allowing more people to leave reviews in different languages.

\section{Discussion}

AI can make people's lives easier in different ways. 
Currently, especially content creation is becoming more efficient, and in the case of writing, tools such as ChatGPT have been revolutionary.
With restaurant reviews, similar tools can help people write more informative and better reviews.
Previous work also highlights that reviews perceived as more informative are often considered more helpful by readers~\cite{luo2021makes,fu2018effects}.
Our findings suggest that users actively seek to produce such informative reviews when given the tools to do so. 
Indeed, users seem to have two primary motivations: they not only seek comprehensive reviews but are also committed to producing detailed content themselves.
Our study also shows a correlation between users' experience levels in online review writing and the richness of detail in their prompts to the AI system (see Section \ref{sec: experience-length}). Although we measure this rather naively with just the word count, it is not far-fetched to suggest that users with greater expertise in writing online reviews typically produce more detailed prompts, indicating that experience can enhance the clarity and specificity with which users express their desires to the AI system. And that, in turn, should lead to people trusting their own abilities in writing. 


\subsection{User Self-efficacy in Providing Useful and Engaging Reviews with AI}
Extensive research has been conducted to assess self-efficacy in writing tasks, and more recently artificial intelligence has been used to improve people's writing skills. For example, in educational contexts, AI chatbots can enhance self-efficacy, learning attitudes, and motivation~\cite{lee2022impacts}. The study indicated that AI chatbots outperform traditional methods in facilitating learning outcomes and emphasize the critical role of AI in boosting self-efficacy, reducing anxiety, and improving writing skills. AI plays an essential role in fostering these improvements~\cite{erito2023exploring}. In addition, it has been shown that undergraduates can learn narrative creation and self-efficacy in writing using similar AI tools~\cite{pellas2023effects}. Furthermore, the potential of AI applications as effective writing assistance tools for non-native post-graduate students in academic English writing has been demonstrated earlier~\cite{nazari2021application}.

Although these studies primarily focus on learning contexts, our research extends this investigation to the domain of online restaurant review writing. As detailed in Section \ref{sec: self-efficiency}, our findings indicate improvement in user self-efficacy when utilising Vocalizer. Participants reported a self-efficacy score of 4.86 out of 10 when writing reviews without AI assistance, indicating substantial room for improvement. 
In the final questionnaire, participants reported a high self-efficacy score of 9.14 out of 10 when using AI-enhanced applications like ours for writing reviews. The quantitative data was additionally supported in our qualitative findings, where users indicated that the AI agent helped them arrange their thoughts and fix their language to something they were much more satisfied with. 

Our prototype is far from a perfect tool, but it acts as a concrete step toward helping people write reviews without perfect command of the English language and with a helpful agent that can structure their thoughts into a sensible review.

\subsection{User Satisfaction and Preferences in AI-Assisted Review Generation}

A study has found that AI-created reviews were less useful, trustworthy and authentic when people were aware that the content was AI-generated, compared to reviews that they knew were written by humans\cite{AMOS2024102163}. \citeauthor{kovacs2024turing} also reported a degree of difficulty in distinguishing human- and AI-generated content ~\cite{kovacs2024turing}. This highlights how the line between human- and AI-generated content is increasingly blurred, pointing to a future where online reviews are increasingly influenced AI technologies.Furthermore, it has been suggested ~\cite{KIM2020101452} that AI-generated content would not differ significantly from human-generated content in terms of perceived quality, readability, and credibility, challenging traditional distinctions between these two types of content.

Our study, in turn, also shows how users had some concerns about the authenticity, or even lost creativity, when using AI in reviews. Although generative AI has shown its capability in producing creative work, humans typically recognise the dangers of overreliance on AI \cite{oppenlaender2023text}. Another way to look at satisfaction is to gauge how willing people are to show their writing publicly, i.e. publish it online. We compared the willingness to share reviews generated by two different versions of Vocalizer. Our analysis revealed a statistically significant difference in willingness to share, with users expressing a markedly higher tendency to share reviews created with the LAV compared to those created with the VOV. 

The general satisfaction with the final review generated by the LAV averaged 6.15 out of 7, reflecting high levels of contentment. Similarly, the perceived usefulness of the AI agent in the review generation process was rated 6.09 out of 7. These results indicate the effectiveness of AI in increasing user satisfaction and engagement with the review application.

\subsection{Limitations \& Future work}
We acknowledge certain limitations. First, the sample size was relatively small, which may limit the generalizability of the findings to a wider population with a more diverse audience and setting.
Prior work has shown that users' personality characteristics impact their perception of LLM-generated advice~\cite{Wester2024PerceptionsAdvice}.
Second, the self-efficacy measurement about the users' prior belief in being able to leave a good review was captured in the final questionnaire, while it may have been more useful to capture it also in the onboarding survey. While this makes it easier for the users to give an assessment as they have other comparison points in mind, we acknowledge there is certain uncertainty in the initial value. Finally, the choice of automatically cleansing the review in the LAV condition could have been done in both versions. This, as we discussed, is a trade-off between offering some automatic AI features to the users of LAV version and creating a more balanced study design between the conditions.

We propose future research avenues. First, researchers could investigate reviews across different cultures, as this would enable offering tips in a culturally personalised way instead of a one-size-fits-all approach. Second, building an entirely personalised review helper agent for the more experienced people based on their past review history is an interesting opportunity to investigate.

\section{Conclusion}
This research has showcased the capability of AI-powered tools to improve user-generated content, particularly regarding online restaurant reviews. Our comparative study of a Voice-Only Version and an LLM-Assisted Version demonstrates that the LAV enhances the quality and coherence of user reviews while also markedly improving user satisfaction and self-efficacy. The ability of the LAV to polish spontaneous input into well-organized content led to increased user engagement and a higher likelihood of publicly sharing reviews, underlining its promise in improving digital content creation. Furthermore, qualitative feedback highlights the potential of LAV in enhancing the review process, although it also raises significant concerns about the authenticity and ethical implications of AI-generated content. These results indicate that with careful integration, advanced AI tools have the potential to enhance the quality of user-created content.

\begin{acks}
This research is partly funded by the Strategic Research Council (SRC), established within the Research Council of Finland (Grants 356128, 335625, 335729), and Research Fellowship funding by the Research Council of Finland (Grants 356128, 349637 and 353790). The research was also partly supported by JSPS Bilateral Collaboration between Japan and Finland (Grant Number: JPJSBP 120232701).
\end{acks}

\bibliographystyle{ACM-Reference-Format}
\bibliography{sample-base}

@article{Talarico2023,
author = {Talarico, Donna},
title = {Engineering answers: Honing your AI prompts for more effective results},
journal = {Recruiting \& Retaining Adult Learners},
volume = {25},
number = {12},
pages = {1-3},
doi = {https://doi.org/10.1002/nsr.31064},
year = {2023}
}

@article{oppenlaender2023text,
  title={Text-to-image generation: Perceptions and realities},
  author={Oppenlaender, Jonas and Visuri, Aku and Paananen, Ville and Linder, Rhema and Silvennoinen, Johanna},
  journal={arXiv preprint arXiv:2303.13530},
  year={2023}
}

@Article{schrepp2017design,
  title={Design and evaluation of a short version of the user experience questionnaire (UEQ-S)},
  author={Schrepp, Martin and Hinderks, Andreas and Thomaschewski, J{\"o}rg},
  journal={International Journal of Interactive Multimedia and Artificial Intelligence, 4 (6), 103-108.},
  year={2017},
  publisher={UNIR}
}

@Article{bandura2006guide,
  title={Guide for constructing self-efficacy scales},
  author={Bandura, Albert and others},
  journal={Self-efficacy beliefs of adolescents},
  volume={5},
  number={1},
  pages={307--337},
  year={2006},
  publisher={Greenwich, CT}
}

@article{pezenka2020factors,
  title={Which factors influence locals’ and visitors’ overall restaurant evaluations?},
  author={Pezenka, Ilona and Weismayer, Christian},
  journal={International Journal of Contemporary Hospitality Management},
  volume={32},
  number={9},
  pages={2793--2812},
  year={2020},
  publisher={Emerald Publishing Limited}
}

@article{LI2019257,
title = {Making restaurant reviews useful and/or enjoyable? The impacts of temporal, explanatory, and sensory cues},
journal = {International Journal of Hospitality Management},
volume = {83},
pages = {257-265},
year = {2019},
issn = {0278-4319},
doi = {https://doi.org/10.1016/j.ijhm.2018.11.002},
author = {Hengyun Li and Chuhan (Renee) Wang and Fang Meng and Zili Zhang},
}

@article{matwick2018restaurant,
  title={Restaurant reviews and college writing: A framework for teaching},
  author={Matwick, Kelsi and Matwick, Keri},
  journal={Teaching Journalism \& Mass Communication},
  volume={8},
  number={2},
  pages={25--37},
  year={2018},
  publisher={Association for Education in Journalism and Mass Communication Small~…}
}

@article{yoo2008motivates,
  title={What motivates consumers to write online travel reviews?},
  author={Yoo, Kyung Hyan and Gretzel, Ulrike},
  journal={Information Technology \& Tourism},
  volume={10},
  number={4},
  pages={283--295},
  year={2008},
  publisher={Cognizant Communication Corporation}
}

@article{NGOYE201447,
title = {The influence of reviewer engagement characteristics on online review helpfulness: A text regression model},
journal = {Decision Support Systems},
volume = {61},
pages = {47-58},
year = {2014},
issn = {0167-9236},
doi = {https://doi.org/10.1016/j.dss.2014.01.011},
author = {Thomas L. Ngo-Ye and Atish P. Sinha},
}

@article{FANG2016498,
    title = {Analysis of the perceived value of online tourism reviews: Influence of readability and reviewer characteristics},
    journal = {Tourism Management},
    volume = {52},
    pages = {498-506},
    year = {2016},
    issn = {0261-5177},
    doi = {https://doi.org/10.1016/j.tourman.2015.07.018},
    author = {Bin Fang and Qiang Ye and Deniz Kucukusta and Rob Law},
}

@article{PARK201567,
title = {Asymmetric effects of online consumer reviews},
journal = {Annals of Tourism Research},
volume = {50},
pages = {67-83},
year = {2015},
issn = {0160-7383},
doi = {https://doi.org/10.1016/j.annals.2014.10.007},
author = {Sangwon Park and Juan L. Nicolau},
}

@article{RACHERLA2012548,
title = {Perceived ‘usefulness’ of online consumer reviews: An exploratory investigation across three services categories},
journal = {Electronic Commerce Research and Applications},
volume = {11},
number = {6},
pages = {548-559},
year = {2012},
note = {Information Services in EC},
issn = {1567-4223},
doi = {https://doi.org/10.1016/j.elerap.2012.06.003},
author = {Pradeep Racherla and Wesley Friske},
}

@article{doi:10.1080/10447318.2020.1812909,
author = {Ni Kang, Ding Ding, M. Birna Van Riemsdijk},
title = {Self-identification with a Virtual Experience and Its Moderating Effect on Self-efficacy and Presence},
journal = {International Journal of Human–Computer Interaction},
volume = {37},
number = {2},
pages = {181--196},
year = {2021},
publisher = {Taylor \& Francis},
doi = {10.1080/10447318.2020.1812909}
}

@article{namkung2008highly,
  title={Are highly satisfied restaurant customers really different? A quality perception perspective},
  author={Namkung, Young and Jang, SooCheong},
  journal={International Journal of Contemporary Hospitality Management},
  volume={20},
  number={2},
  pages={142--155},
  year={2008},
  publisher={Emerald Group Publishing Limited}
}

@article{CHUA2016547,
title = {Helpfulness of user-generated reviews as a function of review sentiment, product type and information quality},
journal = {Computers in Human Behavior},
volume = {54},
pages = {547-554},
year = {2016},
issn = {0747-5632},
doi = {https://doi.org/10.1016/j.chb.2015.08.057},
author = {Alton Y.K. Chua and Snehasish Banerjee},
}

@article{doi:10.1287/mnsc.2022.4359,
author = {Zhang, Mengxia and Luo, Lan},
title = {Can Consumer-Posted Photos Serve as a Leading Indicator of Restaurant Survival? Evidence from Yelp},
journal = {Management Science},
volume = {69},
number = {1},
pages = {25-50},
year = {2023},
doi = {10.1287/mnsc.2022.4359}
}

@article{Liang2019,
  author = {Sai Liang and Markus Schuckert and Rob Law},
  title = {How to improve the stated helpfulness of hotel reviews? A multilevel approach},
  journal = {International Journal of Contemporary Hospitality Management},
  volume = {31},
  number = {2},
  pages = {953-977},
  year = {2019},
  doi = {10.1108/IJCHM-02-2018-0134},
}

@article{mudambi2010,
  author = {Mudambi, Sriram and Schuff, David A.},
  title = {What Makes a Helpful Online Review? A Study of Customer Reviews on Amazon.comCOM},
  journal = {Journal of Marketing Research},
  volume = {47},
  number = {1},
  pages = {133--147},
  year = {2010},
  publisher = {American Marketing Association}
}

@article{alfira2024enhancing,
  title={Enhancing Consumer Trust and Purchase Decisions Through Online Reviews and Hedonic Factors: A Study of Jakarta Coffee Shops},
  author={Alfira, Hana and Raharjo, Pamuji Gesang and Kristiadi, Al Agus},
  journal={Valley International Journal Digital Library},
  pages={6804--6814},
  year={2024}
}

@article{haro2024influence,
  title={Influence of electronic word-of-mouth on restaurant choice decisions: does it depend on gender in the millennial generation?},
  author={Haro-Sosa, Giovanny and Moliner-Vel{\'a}zquez, Beatriz and Gil-Saura, Irene and Fuentes-Blasco, Mar{\'\i}a},
  journal={Journal of Theoretical and Applied Electronic Commerce Research},
  volume={19},
  number={1},
  pages={615--632},
  year={2024},
  publisher={MDPI}
}

@inproceedings{singh2023word,
  title={E-Word of Mouth and Restaurant-goers: an Empirical Study on the Influence of e-WOM on the Selection of Restaurants},
  author={Singh, Meghna and Fatma, Anam and Bhatt, Vimal and Sinha, Binod},
  booktitle={2023 International Conference On Cyber Management And Engineering (CyMaEn)},
  pages={474--478},
  year={2023},
  organization={IEEE}
}

@article{Saddam2024,
author = {Saddam Abdullah and Philippe Van Cauwenberge and Heidi Vander Bauwhede and Peter O’Connor},
title ={Review Ratings, Sentiment in Review Comments, and Restaurant Profitability: Firm-Level Evidence},

journal = {Cornell Hospitality Quarterly},
volume = {65},
number = {3},
pages = {378-392},
year = {2024},
doi = {10.1177/19389655231214758}
}

@article{LI2023104707,
title = {Restaurant survival prediction using customer-generated content: An aspect-based sentiment analysis of online reviews},
journal = {Tourism Management},
volume = {96},
pages = {104707},
year = {2023},
issn = {0261-5177},
doi = {https://doi.org/10.1016/j.tourman.2022.104707},
author = {Hengyun Li and Bruce X.B. Yu and Gang Li and Huicai Gao}
}

@inproceedings{Fernandes2018,
author = {Fernandes, Michael and Walls, Logan and Munson, Sean and Hullman, Jessica and Kay, Matthew},
title = {Uncertainty Displays Using Quantile Dotplots or CDFs Improve Transit Decision-Making},
year = {2018},
isbn = {9781450356206},
publisher = {Association for Computing Machinery},
address = {New York, NY, USA},
doi = {10.1145/3173574.3173718},
booktitle = {Proceedings of the 2018 CHI Conference on Human Factors in Computing Systems},
pages = {1–12},
numpages = {12},
location = {Montreal QC, Canada},
series = {CHI '18}
}

@misc{lee2024,
      title={Navigating the Path of Writing: Outline-guided Text Generation with Large Language Models}, 
      author={Yukyung Lee and Soonwon Ka and Bokyung Son and Pilsung Kang and Jaewook Kang},
      year={2024},
      eprint={2404.13919},
      archivePrefix={arXiv},
      primaryClass={cs.CL},
      url={https://arxiv.org/abs/2404.13919}, 
}

@article{zhao2023investigating,
  title={Investigating Table-to-Text Generation Capabilities of LLMs in Real-World Information Seeking Scenarios},
  author={Zhao, Yilun and Zhang, Haowei and Si, Shengyun and Nan, Linyong and Tang, Xiangru and Cohan, Arman},
  journal={arXiv preprint arXiv:2305.14987},
  year={2023}
}

@misc{marco2024pronvspromptlarge,
      title={Pron vs Prompt: Can Large Language Models already Challenge a World-Class Fiction Author at Creative Text Writing?}, 
      author={Guillermo Marco and Julio Gonzalo and Ramón del Castillo and María Teresa Mateo Girona},
      year={2024},
      eprint={2407.01119},
      archivePrefix={arXiv},
      primaryClass={cs.CL},
      url={https://arxiv.org/abs/2407.01119}, 
}

@misc{brightlocal2023,
  author = {BrightLocal},
  title = {Local Consumer Review Survey},
  year = {2023},
  url = {https://www.brightlocal.com/research/local-consumer-review-survey/},
  note = {[Accessed: 11-Aug-2024]}
}

@article{sourabh_jain_2024,
        title={ Influence of Online Consumer Reviews on Brand Reputation },
        author={ Sourabh Jain },
        journal={ International Journal For Science Technology And Engineering },
        year={ 2024 },
        publisher={ International Journal for Research in Applied Science and Engineering Technology (IJRASET) },
        volume={ 12 },
        pages={ 4785-4790 },
        number={ 4 },
        
        
        doi={ 10.22214/ijraset.2024.61013 },  
      }

@inproceedings{Lin2024,
author = {Lin, Susan and Warner, Jeremy and Zamfirescu-Pereira, J.D. and Lee, Matthew G and Jain, Sauhard and Cai, Shanqing and Lertvittayakumjorn, Piyawat and Huang, Michael Xuelin and Zhai, Shumin and Hartmann, Bjoern and Liu, Can},
title = {Rambler: Supporting Writing With Speech via LLM-Assisted Gist Manipulation},
year = {2024},
isbn = {9798400703300},
publisher = {Association for Computing Machinery},
address = {New York, NY, USA},
doi = {10.1145/3613904.3642217},
booktitle = {Proceedings of the CHI Conference on Human Factors in Computing Systems},
articleno = {1043},
numpages = {19},
location = {Honolulu, HI, USA},
series = {CHI '24}
}

@article{luo2021makes,
  title={What makes a helpful online review? Empirical evidence on the effects of review and reviewer characteristics},
  author={Luo, Lijuan and Duan, Siqi and Shang, Shanshan and Pan, Yu},
  journal={Online Information Review},
  volume={45},
  number={3},
  pages={614--632},
  year={2021},
  publisher={Emerald Publishing Limited}
}

@article{fu2018effects,
  title={Effects of membership tier on user content generation behaviors: Evidence from online reviews},
  author={Fu, Dongpu and Hong, Yili and Wang, Kanliang and Fan, Weiguo},
  journal={Electronic Commerce Research},
  volume={18},
  pages={457--483},
  year={2018},
  publisher={Springer}
}

@article{lee2022impacts,
  title={Impacts of an AI-based cha bot on college students’ after-class review, academic performance, self-efficacy, learning attitude, and motivation},
  author={Lee, Yen-Fen and Hwang, Gwo-Jen and Chen, Pei-Ying},
  journal={Educational technology research and development},
  volume={70},
  number={5},
  pages={1843--1865},
  year={2022},
  publisher={Springer}
}

@article{erito2023exploring,
  title={Exploring ESP Students’ Perception toward the Potential of Artificial Intelligence to Promote Students’ Self-Efficacy in English Writing Skill},
  author={Erito, Silvia Nanda Putri},
  journal={Journal of English Language Learning},
  volume={7},
  number={2},
  pages={457--464},
  year={2023}
}

@article{pellas2023effects,
  title={The Effects of Generative AI Platforms on Undergraduates’ Narrative Intelligence and Writing Self-Efficacy},
  author={Pellas, Nikolaos},
  journal={Education Sciences},
  volume={13},
  number={11},
  pages={1155},
  year={2023},
  publisher={MDPI}
}

@article{nazari2021application,
  title={Application of Artificial Intelligence powered digital writing assistant in higher education: randomized controlled trial},
  author={Nazari, Nabi and Shabbir, Muhammad Salman and Setiawan, Roy},
  journal={Heliyon},
  volume={7},
  number={5},
  year={2021},
  publisher={Elsevier}
}

@article{AMOS2024102163,
    title = {Consumer reactions to perceived undisclosed ChatGPT usage in an online review context},
    journal = {Telematics and Informatics},
    volume = {93},
    pages = {102163},
    year = {2024},
    issn = {0736-5853},
    doi = {https://doi.org/10.1016/j.tele.2024.102163},
    author = {Clinton Amos and Lixuan Zhang}
}

@article{kovacs2024turing,
  title={The Turing test of online reviews: Can we tell the difference between human-written and GPT-4-written online reviews?},
  author={Kov{\'a}cs, Bal{\'a}zs},
  journal={Marketing Letters},
  pages={1--16},
  year={2024},
  publisher={Springer}
}

@article{KIM2020101452,
title = {Can AI be a content generator? Effects of content generators and information delivery methods on the psychology of content consumers},
journal = {Telematics and Informatics},
volume = {55},
pages = {101452},
year = {2020},
issn = {0736-5853},
doi = {https://doi.org/10.1016/j.tele.2020.101452},
author = {Jina Kim and Soyeon Shin and Kunwoo Bae and Soyoung Oh and Eunil Park and Angel P. {del Pobil}}
}

@article{HUANG2018430,
title = {Online customer reviews and consumer evaluation: The role of review font},
journal = {Information \& Management},
volume = {55},
number = {4},
pages = {430-440},
year = {2018},
issn = {0378-7206},
doi = {https://doi.org/10.1016/j.im.2017.10.003},
author = {Yunhui Huang and Changxin Li and Jiang Wu and Zhijie Lin}
}

@inproceedings{Suzuki2020,
author = {Suzuki, Rinji and Akiyama, Kazuhiro and Kumamoto, Tadahiko and Nadamoto, Akiyo},
title = {Analysis of High-value Reviews based on Sentiment},
year = {2020},
isbn = {9781450371797},
publisher = {Association for Computing Machinery},
address = {New York, NY, USA},
doi = {10.1145/3366030.3366038},
booktitle = {Proceedings of the 21st International Conference on Information Integration and Web-Based Applications \& Services},
pages = {232–239},
numpages = {8},
location = {Munich, Germany},
series = {iiWAS2019}
}

@article{tandon2021analyzing,
  title={Analyzing the impact of review recency on helpfulness through econometric modeling},
  author={Tandon, Abhishek and Aakash, Aakash and Aggarwal, Anu G and Kapur, PK},
  journal={International Journal of System Assurance Engineering and Management},
  volume={12},
  pages={104--111},
  year={2021},
  publisher={Springer}
}

@article{ahn2024impact,
  title={The Impact of Online Reviews on Consumers’ Purchase Intentions: Examining the Social Influence of Online Reviews, Group Similarity, and Self-Construal},
  author={Ahn, Yunjeong and Lee, Jieun},
  journal={Journal of Theoretical and Applied Electronic Commerce Research},
  volume={19},
  number={2},
  pages={1060--1078},
  year={2024},
  publisher={MDPI}
}

@article{braunthematic,
author = {Virginia Braun and Victoria Clarke},
title = {Using thematic analysis in psychology},
journal = {Qualitative Research in Psychology},
volume = {3},
number = {2},
pages = {77--101},
year = {2006},
publisher = {Routledge},
doi = {10.1191/1478088706qp063oa}
}

@article{fang2023systematic,
  title={A systematic review of artificial intelligence technologies used for story writing},
  author={Fang, Xiaoxuan and Ng, Davy Tsz Kit and Leung, Jac Ka Lok and Chu, Samuel Kai Wah},
  journal={Education and Information Technologies},
  volume={28},
  number={11},
  pages={14361--14397},
  year={2023},
  publisher={Springer}
}

@inproceedings{li2024value,
  title={The Value, Benefits, and Concerns of Generative AI-Powered Assistance in Writing},
  author={Li, Zhuoyan and Liang, Chen and Peng, Jing and Yin, Ming},
  booktitle={Proceedings of the CHI Conference on Human Factors in Computing Systems},
  pages={1--25},
  year={2024}
}

@article{gambetti2023combat,
  title={Combat ai with ai: Counteract machine-generated fake restaurant reviews on social media},
  author={Gambetti, Alessandro and Han, Qiwei},
  journal={arXiv preprint arXiv:2302.07731},
  year={2023}
}

@article{jayavalan2018effectiveness,
  title={Effectiveness of online grammar checker to improve secondary students’ English narrative essay writing},
  author={Jayavalan, Kalpana and Razali, Abu Bakar and Abu, B},
  journal={International Research Journal of Education and Sciences},
  volume={2},
  number={1},
  pages={1--6},
  year={2018}
}

@article{shahriar2023potential,
  title={Potential Success in English Writing Skills Using Artificial Intelligence “Grammarly”},
  author={Shahriar, Arafat and Laboni, Fatema Begum},
  journal={DIU Journal of Humanities and Social Science},
  volume={10},
  number={1},
  pages={57--75},
  year={2023}
}

@inproceedings{huang2020effectiveness,
  title={The effectiveness of using grammarly to improve students' writing skills},
  author={Huang, Hui-Wen and Li, Zehui and Taylor, Linda},
  booktitle={Proceedings of the 5th International Conference on Distance Education and Learning},
  pages={122--127},
  year={2020}
}

@article{rice2007using,
  title={Using customer reviews to build critical reading skills},
  author={Rice, Mary},
  journal={English Journal},
  volume={97},
  number={1},
  pages={89--93},
  year={2007},
  publisher={ncte. org}
}

@article{furner2021,
author = {Furner, Christopher and Yoon, Tom and Zinko, Robert and Goh, Samuel},
year = {2021},
month = {07},
pages = {1-15},
title = {The Influence of Reviewer and Consumer Congruence in Online Word-of-Mouth Transactions},
volume = {19},
journal = {Journal of Electronic Commerce in Organizations},
doi = {10.4018/JECO.2021070101}
}

@inproceedings{bartoli2016best,
  title={" Best dinner ever!!!": automatic generation of restaurant reviews with LSTM-RNN},
  author={Bartoli, Alberto and De Lorenzo, Andrea and Medvet, Eric and Morello, Dennis and Tarlao, Fabiano},
  booktitle={2016 IEEE/WIC/ACM International Conference on Web Intelligence (WI)},
  pages={721--724},
  year={2016},
  organization={IEEE}
}

@inproceedings{gambetti2024aigen,
  title={AiGen-FoodReview: A Multimodal Dataset of Machine-Generated Restaurant Reviews and Images on Social Media},
  author={Gambetti, Alessandro and Han, Qiwei},
  booktitle={Proceedings of the International AAAI Conference on Web and Social Media},
  volume={18},
  pages={1935--1945},
  year={2024}
}

@article{Miriam2020,
author = {Miriam van Mersbergen, Lisa A. Vinney and Alexis E. Payne},
title = {Cognitive influences on perceived phonatory exertion using the Borg CR10},
journal = {Logopedics Phoniatrics Vocology},
volume = {45},
number = {3},
pages = {123--133},
year = {2020},
publisher = {Taylor \& Francis},
doi = {10.1080/14015439.2019.1617895},

    note ={PMID: 31190588},


URL = { 
    
        https://doi.org/10.1080/14015439.2019.1617895
    
    

},
eprint = { 
    
        https://doi.org/10.1080/14015439.2019.1617895
    
    

}

}

@article{Wester2024PerceptionsAdvice,
title = {Exploring people's perceptions of LLM-generated advice},
journal = {Computers in Human Behavior: Artificial Humans},
volume = {2},
number = {2},
pages = {100072},
year = {2024},
issn = {2949-8821},
doi = {10.1016/j.chbah.2024.100072},
author = {Joel Wester and Sander {de Jong} and Henning Pohl and Niels {van Berkel}},
}

\newpage
\appendix

\section{Appendix A}\label{sec:appendix_a}
This appendix provides the detailed instructions used to guide the Large Language Model (LLM) in improving user voice transcribe reviews. 

\begin{lstlisting}[language=html, caption=Initial Review Improvement LLM Prompt, label=lst:ai-prompts]
Imagine yourself as a human. You should characterize the review as human and with human vibes. You should not generate the review as an AI. This is a must. 
You should not add any words like `Here is the refined review', `Here is the improved review,' or any other words at the beginning of the refined review. Just refine the review and send it back. This is a must.
Please refine this review by removing unnecessary rambling and ensuring coherence.
Focus on eliminating filler words like `um' and `ah' while maintaining the original tone.
Keep the content and tone as close to the original audio review as possible.
Ensure that the tweaks maintain the authenticity of the reviewer's voice.
The readability of the review is the most important thing. Please ensure that the refined review is easy to read and understand. This is a must. 
The review should remain true to the original sentiment expressed in the input review.
Please don't add any additional words to make it creative or more informative. Just refine the review and send it back. 
Analyze the English level of the review and make your English level similar to the review. As an example, if the review is simple, make your English simple. If the review is complex, make your English complex. If the review is formal, make your English formal. If the review is informal, make your English informal. If the review is casual, make your English casual. If the review is professional, make your English professional.
Please ensure your generated review English level is similar to the original review English level. First, identify the English level of review and generate according to it. Most of the time, users give reviews in simple English. if it is simple English, don't use any fancy English words, it should be simple English. This is a must must.
I gave you the name of the restaurant. You dont need to add that in the review. Sometimes you might assume another name as the name of a restaurant. That is why I added it. If the user said the restaurant name in the review, you can add the value of the restaurant name. Maybe in the text spelling might be wrong because the review is a voice transcription. Otherwise, do not add it. This is a must
If you cannot generate results or there is not enough information, please send a message like `Sorry, we cannot improve this review' to the user.
The readability of a review text is correlated with the perceived helpfulness of reviews. so make sure the refined review is easy to read and understand. this is a must
\end{lstlisting}

\section{Appendix B}\label{sec:appendix_b}
This Appendix details the instructions to refine the initial improved reviews using an AI agent. 

\begin{lstlisting}[language=html, caption=AI Agent Prompts, label=lst:ai-agent-prompts]
We have an application that collects user audio reviews and then transcribes and sends them to LLM to remove filler words, making them more coherent. We will display the results on the results display page. Your task is to refine the review based on the user's instructions
Please refine the generatedText based on the following instructions by user refineInstructions
Please keep the English level the same as the original unless users request to change it by refineInstructions
Please don't add words like `Here's the refined review' or `Here's the improved review' or any other words at the beginning of the refined review. Just refine the review and send it back.
If the user asks to refine only one part of the review, like rewriting only one incident, you need to refine only that incident. You don't need to refine the whole review. keep the other parts as previously and only change the part that the user asks for. You should send the whole review back, but change the only sentence that the user asked. This is a super must must.
The readability of a review text is correlated with the perceived helpfulness of the reviews. So make sure the refined review is easy to read and understand. And also English should be simple that everyone can understand. This is a must.
Please ensure the refined review maintains the original sentiment and tone unless the user requests to change it by refineInstructions.
If you cannot generate results, send a message like `Sorry, we cannot refine this review' to the user. this is a must
\end{lstlisting}

\section{Appendix C}\label{sec:appendix_c}
This appendix outlines the review feedback tips to improve the quality and usefulness of user reviews, as generated by the language model. The prompts provided are informed by recent research ~\cite{pezenka2020factors, LI2019257, matwick2018restaurant, yoo2008motivates, NGOYE201447, FANG2016498, PARK201567, RACHERLA2012548, doi:10.1080/10447318.2020.1812909, namkung2008highly, CHUA2016547} on the characteristics of highly valuable reviews.

\begin{lstlisting}[language=html, caption=Review Improvement Tips LLM Prompt, label=lst:suggestion_prompts]
Don't tell users to rate the review on a scale when you give tips. We already have a star rating system. We are evaluating the overall rating for food, service, and atmosphere on a scale of 1 to 5 from the user. So we do not need to tell them again to do that in the review. Please keep this in mind when writing a review. Do not print this in a genarated review, this is for your guidance. This is a must.
If you cannot generate results, send a message like `Sorry, we don't have any improvement tips for this review' to the user. And don't add any other words to that message. Send what I typed here. This is a must.
You must focus on giving tips to make the review high quality and valuable for other consumers. I will give you some latest research on what a good review looks like. Please, follow those findings. This is a must.
The researchers found that the most helpful reviews are those that are specific and detailed. Therefore, when giving tips, focus on making the review more specific and detailed.
The researchers found that if there is an indication of the reason or cause-and-effect relationship in the review, it is more helpful to other consumers.
Researchers found that food has the strongest impact on the overall evaluation of the restaurant.
Researchers found that reviews with more explanatory cues are more likely to be perceived as useful and enjoyable
Researchers found that in-depth reviews to explain why he/she liked or disliked the products or services, rather than simply posting a positive or negative review without further explanations, are helpful for other consumers.
The researchers found that restaurants and online review platforms should encourage review writers to provide reaction words that communicate feelings, emotions, and subjective evaluations, as well as experiences in their reviews.
The researchers found that the readability of a review text is correlated with the perceived helpfulness of the reviews
Researchers found that reviews with precise or easy-to-understand writing styles will receive more helpfulness votes.
The researchers found that reviews expressing extreme sentiment would be considered valuable.
\end{lstlisting}

\onecolumn
\clearpage
\newpage
\section{Appendix E}\label{sec:appendix_e}
This appendix provides several examples of AI-generated improvement tips for users' reviews. Table \ref{tab: ai-tips} presents examples of users' initial reviews alongside \textit{AI improvement tips} generated by a language model.

\begin{table}[H]
    \centering
    \caption{Examples of AI improvement tips in the LAV condition}
    \scriptsize{
    \begin{tabular}{p{3.5cm}p{13.6cm}}
    \toprule
    \textbf{Initial improved Review} & \textbf{AI improvement tips}  \\
    \midrule
    The minced meat dish was enjoyable, served with salad. However, the flavor of the meat was a bit mild for my taste. & Here are some tips on improving the quality of your review: 
    
    1. Be More Specific: You could provide more details about the minced meat dish such as its texture, the type of meat used, and how it's prepared. For example, "The minced beef in the dish was soft, finely chopped, and seemed to be cooked with minimal spices." 
    
    2. Cause-effect Relationships: Specify what about the flavor was mild and how it affected your experience. For instance, you might say, "The flavor of the meat was milder than I prefer, lacking in spices which left it tasting a bit bland." 
    
    3. Details about salad: Adding more details about the salad can also improve your review. You might describe the ingredients in the salad, whether it was fresh, and how well it complemented the meat dish. 
    
    4. Reaction Words: Use more expressive language to convey your feelings towards the dish. This gives a more personal touch to your review and makes it more enjoyable for readers. You could say something like "I was really looking forward to a burst of flavors but was slightly disappointed by the milder than expected taste". 
    
    5. Readability: Try to use shorter sentences and simpler words to ensure your review is easy to understand for all readers. 
    
    6. Extreme Sentiment: Conveying a stronger sentiment can make your review more valuable. For example, you could say, "While the meat dish was decent, I couldn't help but wish for a more robust flavor to satiate my taste buds." Remember, these are just tips to guide you in writing a more detailed and helpful review! \\
    \midrule
    I frequently dine at this restaurant due to its extensive variety of dishes. The flexibility in serving options makes it my go-to choice, especially when I'm at the university. & This review is quite good, but there are ways to improve. Here are some suggestions: 
    
    1. Include more specifics: Your review mentions an "extensive variety of dishes" and flexibility in the serving options. Could you mention a few specific dishes that you enjoyed or unique serving options that stand out? This would bolster your review's credibility. 
    
    2. Add more details about the food: Considering food plays a pivotal role in a restaurant's evaluation, it would add value if you could provide more information about the food quality, taste or presentation. 
    
    3. Reaction words and experiences: It can be quite helpful to emphasize how the restaurant's dishes made you feel or enhanced your dining experience. Did it exceed your expectations or give you a memorable moment? 
    
    4. Cause-and-effect explanations: You mentioned dining at the restaurant when you're at the university. Could you explain why? For example, is it conveniently located? Or is it a perfect spot for a quick lunch between classes? Remember, the more precise and easy to understand your review, the more helpful it will be to others. \\
    \midrule
    I visited this place and was pleasantly surprised by its ambiance. The food was particularly delicious. It's not your typical lunch spot, but if you're looking for a meal during breakfast, lunch, or even a break, it's worth a visit. They usually close around 7 or 8, providing ample time for those who work late, like at the university. It's a great option as most other restaurants tend to close earlier.  & The review you've provided has many strong points, but could benefit from additional detail and specificity to enhance its value for other consumers. Here are few suggestions: 
    
    1. Provide specific examples: What exactly about the ambiance pleasantly surprised you? Was it the decor, lighting, music, or seating arrangement? As for the food, share which dishes you particularly enjoyed and why. 
    
    2. Include the service aspect: Mention the quality of service received. Was the staff friendly, courteous, and prompt in serving you? 
    
    3. Write about your personal experience: Adding personal anecdotes or stories will make your review more relatable and interesting. Maybe you could share your reasons for visiting this place. 
    
    4. Communicate your emotions: Express more clearly how the restaurant made you feel. Happy, satisfied, delighted? This will add emotional depth to your review. 5. Improve readability: Verify that your review is easy-to-read and follows a coherent order. Ensure your sentences are not too long and your ideas flow naturally. Remember, a great review is more than an evaluation, it's a story of your experience. Happy reviewing! \\
    \bottomrule
    \end{tabular}
    }    
    \label{tab: ai-tips}
\end{table}

\section{Appendix D}\label{sec:appendix_d}
This appendix provides detailed examples of how users have requested refinements to their reviews. In the results section, Table \ref{tab: user-promts} shows various types of modification. In table \ref{tab: Prompts2}, an example is included for each type of modification. The \textit{Original Review} represents the initial improved review, while the \textit{Cleanup Review} showcases how the AI agent refined the review according to the user's instructions in the \textit{Instructions} mentioned in the table.

\begin{table}[H]
    \centering
    \caption{Examples of how Users Wanted their Reviews Edited in the LAV condition}
    \scriptsize{
    \begin{tabular}{p{6.8cm}p{6.8cm}p{1.8cm}p{0.8cm}}
    \toprule
    \textbf{Original Review} & \textbf{Cleanup Review} & \textbf{Instructions} & \\
    \midrule
    The salad options today were notably enjoyable, with cooked Brussels sprouts and red onions included & The salad options today were good, especially the cooked Brussels sprouts and red onions. The fish spiced with lemon was also very good. & fish spiced with lemon was very good & Add details\\
    \midrule
    The lunch today offered many tasty choices, which made picking one slightly difficult. I had to decide between breaded steak and breaded bean and vegetable bricks or balls. In the end, I chose the breaded steak with mashed potatoes, and it was really good. This meal made my day. However, the salad wasn't as fresh as it normally is.& The lunch today had many delicious options, which made choosing one a bit hard. I had to pick between breaded steak and breaded bean and vegetable croquettes. Finally, I went for the breaded steak with mashed potatoes, and it was really good. It made my day. But the salad wasn't as fresh as usual.& not bricks or balls but maybe crotchets & Clarify Ambiguity \\
    \midrule
    For today's lunch, we had a selection of petite steaks, a variety of salads, standard potatoes, and pasta. I opted for the salad and steaks. However, I was slightly disappointed as there were only two steaks per person, which I found to be a small portion. On the bright side, the salad was enjoyable, the vegetables were fresh, and overall, the lunch felt well-rounded. The spreads and bread options also added a nice touch to the meal.& For today's lunch, we had petite steaks, a variety of salads, standard potatoes, and pasta. I chose the salad and steaks. But, I was a bit let down because there were only two steaks for each person, which to me, seemed like a small amount. On the plus side, the salad was good, the veggies were fresh, and all in all, the lunch was balanced. The spreads and different types of bread also made the meal better.& Take away the words ``selection of'' from the final text & Omit Information\\
    \midrule
    This place can be busy, showing people like it. The seats aren't very comfortable, but you can find better seats near the university entrance. The food is simple but has fresh vegetables, like ******'s standard. They use fresh and healthy ingredients that make the food taste good. Usually, you don't have to wait long for your food. I often visit this place when other campus restaurants are closed. The bread, cheese, and soda they serve act as comfort food for us students who work hard for long hours.& This place can be busy, which shows people like it. The seats aren't very comfortable, but there are better seats near the university entrance. The food is simple, but they use fresh vegetables, similar to ******'s standard. They use fresh and healthy ingredients that make the food taste good. Usually, you don't have to wait long for your food. I often visit this place when other campus restaurants are closed. The sandwiches they serve act as a comfort food for us students who work hard for long hours.& replace bread cheese soda with sandwiches & Correct Information\\
    \midrule
    This spot is not just great for coffee breaks with a range of breads to pick from, but their desserts and fruits are also very good. They could add more items to their salad menu to offer more lunch options with vegetables. The people who work here are nice, which makes the visit pleasant. The place feels welcoming with its dark wooden tables and views outside. They have seating outdoors during the late spring, summer, and early fall. However, you should know that the ground outside isn't flat, which may cause the tables and chairs to be shaky. & This place should definitely add more lunch options, especially with vegetables. Right now, they've got a salad menu, but it needs more items. They also have a variety of breads that are perfect for quick coffee breaks. And don't miss out on their desserts and fruits - they're good too. When you visit, you'll find the staff is friendly which makes your stay enjoyable. The feel of the place is inviting with its dark wooden tables and the outside view. In the late spring, summer, and early fall, they have seating outside. But be aware, the ground outside isn't level, so the tables and chairs might be a bit shaky. & tell first about lunch options. then salad and bread. then desserts. last about outside & Focus Adjustment\\
    \midrule
    My recent visit to this restaurant was quite disappointing. I was under the impression that I could help myself to as much food as I wanted, as it was advertised. However, when I tried to take a generous serving of pasta, a staff member stopped me. This was surprising and unpleasant, especially since it's supposed to be a buffet. Despite this incident, I generally enjoy the food here and visit on a weekly basis. & I wasn't too satisfied with my last visit to this restaurant. I thought I could have as much food as I liked because that's what they said. But a staff member didn't let me when I tried to get a lot of pasta. I wasn't happy about this because it's a buffet. But overall, I like the food here and come every week.& convert this review like google review & Platform Adaptation\\
    \hline
   I visited this place and was pleasantly surprised by its ambience. The food was particularly delicious. It's not your typical lunch spot, but if you're looking for a meal during breakfast, lunch, or even a break, it's worth a visit. They usually close around 7 or 8, providing ample time for those who work late, like at the university. It's a great option as most other restaurants tend to close earlier. & I dropped by this spot and boy, was I pleased with the vibe it had going! The food was super tasty. It's not just your run-of-the-mill lunch place, you know? If you're up for a bite during breakfast, lunch, or need a food pit-stop, give this one a go. They're usually open till about 7 or 8, works perfect if you're stuck in late at work, like at the university. What a blessing, especially when most places shut their doors a lot earlier. & Make it more friendly   & Formality Decrease\\

    \midrule
    The food served was enjoyable, although the portion size could have been slightly larger. Nevertheless, the overall experience was positive due to the delicious taste of the dishes. & The meal we had was quite delightful, though it would have been even better if we had a bit more on our plates. Still, the flavour of the dishes made our experience quite pleasant overall.  & Be more polite & Formality Increase\\
    \midrule
    The dining experience was enjoyable, with delicious food that was quickly served, catering well to those in a rush. However, the inclusion of international staff could greatly enhance communication, particularly for ******* speakers who may struggle with understanding different accents. & Eating here was a pleasure, with tasty food served promptly, which is excellent for those on a tight schedule. A suggestion to improve the service further could be to consider including staff who speak different languages, especially *******. This might ease the communication for those who find different accents challenging to comprehend. & don't sound like complaining, it's just an advice & Positivity Increase\\
    \midrule
    I frequently dine at this restaurant due to its extensive variety of dishes. The flexibility in serving options makes it my go-to choice, especially when I'm at the university. & I often eat at this restaurant because it's got loads of dishes. The way they serve food is the only reason I go there, especially when I'm stuck at the university. & Make it rude & Positivity Decrease\\
    \midrule
    I tried the small meatballs yesterday, and they were really tasty with a rosemary sauce. But the potatoes were a bit overcooked for my taste. The salads were nice, and I enjoyed that they kept giving us yummy croutons. They even offered different types of bread spreads, like herbal and garlic, not just the normal ones. Even though I wasn't a fan of the potatoes, I had a fun time there, so I'd rate it 4 stars out of 5. & I gave the tiny meatballs a try yesterday, and boy were they delicious! They had this rosemary sauce that was just to die for. However, the potatoes were a little too cooked for my liking. The salads? Absolutely fantastic! I loved how they kept bringing us these tasty croutons that I just couldn't get enough of. They even had a variety of bread spreads, including herbal and garlic, not just your regular ones. Sure, the potatoes weren't my favourite, but I still had such a great time there! I'd definitely give it a solid 4 stars out of 5. & Make it sound more excited & Increase Excitement\\
    \midrule 
    The meal was exceptionally good. The skewered meat was full of flavor and the salads served today caught my taste. The bread, too, was incredibly delicious. The food quality was outstanding today with a wide variety of choices. Thanks to ****. & The meal was good. The skewered meat had flavor and the salads served today were to my taste. The bread was also delicious. The food quality was good today and there were many choices. Thanks to ****. & A bit less of excitement  & Decrease Excitement\\
    \midrule
    ***** is a great choice for healthy food with a wide variety of toppings to choose from. It's usually open, making it a convenient option for late-night meals. The pricing is reasonable, and I found the submarine sandwich to be particularly affordable. & ***** offers healthy food with many toppings. It is often open, even late at night. The prices are fair, especially for the submarine sandwich. & Make it more precise or easy to understand  & Readability Improvement\\
    \midrule
    The cuisine offered is both delicious and wholesome, although it leans towards the pricier side. For those seeking a similar experience on a budget, Cafe Hub is a commendable alternative. Otherwise, ******* comes highly recommended. & The food's yummy and good for you, but it's a bit expensive. If you're looking for something cheaper, try Cafe Hub. But still, ******* is worth a go. & make it sound casual and short & Length Adjustment\\
    \bottomrule
    \end{tabular}
    }    
    \label{tab: Prompts2}
\end{table}

\end{document}